\documentclass[11pt,a4paper]{article}

\usepackage{natbib}
\usepackage{anysize}
\usepackage{color}
\usepackage{amsmath,amssymb,amsthm,amsfonts,amsopn,bm,latexsym}
\usepackage{verbatim,graphicx}
\usepackage[latin1]{inputenc}
\usepackage[english]{babel}

\marginsize{3cm}{3cm}{3cm}{3cm}

\DeclareMathOperator{\E}{\boldsymbol{\mathbb{E}}}
\DeclareMathOperator{\Var}{Var}
\DeclareMathOperator{\Cov}{Cov}

\DeclareMathOperator{\MSE}{MSE}
\DeclareMathOperator{\MISE}{MISE}

\theoremstyle{plain}
\newtheorem{theorem}{Theorem}
\newtheorem{lemma}{Lemma}
\theoremstyle{remark}
\newtheorem{remark}{Remark}

\makeatletter
\def\widebreve{\mathpalette\wide@breve}
\def\wide@breve#1#2{\sbox\z@{$#1#2$}%
     \mathop{\vbox{\m@th\ialign{##\crcr
\kern0.08em\brevefill#1{0.8\wd\z@}\crcr\noalign{\nointerlineskip}%
                    $\hss#1#2\hss$\crcr}}}\limits}
\def\brevefill#1#2{$\m@th\sbox\tw@{$#1($}%
  \hss\resizebox{#2}{\wd\tw@}{\rotatebox[origin=c]{90}{\upshape(}}\hss$}
\makeatletter

\makeatletter
\def\widecheck{\mathpalette\wide@check}
\def\wide@check#1#2{\sbox\z@{$#1#2$}%
     \mathop{\vbox{\m@th\ialign{##\crcr
\kern0.08em\checkfill#1{0.8\wd\z@}\crcr\noalign{\nointerlineskip}%
                    $\hss#1#2\hss$\crcr}}}\limits}
\def\checkfill#1#2{$\m@th\sbox\tw@{$#1($}%
  \hss\resizebox{#2}{\wd\tw@}{\rotatebox[origin=c]{90}{\upshape$\langle$}}\hss$}
\makeatletter

\title{Estimation of density functionals via cross-validation}

\author{Jos\'e E. Chac\'on\footnote{IMUEx, Departamento de
		Matem\'aticas, Universidad de Extremadura, Badajoz, Spain. E-mail:
		{\tt jechacon@unex.es}} \ and Carlos Tenreiro\footnote{CMUC, Department of Mathematics, University of Coimbra, Coimbra, Portugal. Email: {\tt tenreiro@mat.uc.pt}}}

\begin{document}
	
	\maketitle
	
	\begin{abstract}
		\noindent In density estimation, the mean integrated squared error (MISE) is commonly used as a measure of performance. In that setting, the cross-validation criterion provides an unbiased estimator of the MISE minus the integral of the squared density. Since the minimum MISE is known to converge to zero, this suggests that the minimum value of the cross-validation criterion could be regarded as an estimator of minus the integrated squared density. This novel proposal presents the outstanding feature that, unlike all other existing estimators, it does not need the choice of any tuning parameter. Indeed, it is proved here that this approach results in a consistent and efficient estimator, with remarkable performance in practice. Moreover, apart from this base case, it is shown how several other problems on density functional estimation can be similarly handled using this new principle, thus demonstrating full potential for further applications.
	\end{abstract}
	
	\medskip
	\noindent {\it Keywords:} cross-validation, density functional, kernel smoothing, tuning parameter selection

	\newpage
	
	\section{Introduction}
	
	Cross-validation is a very general and widely used technique in statistics and machine learning \citep[see][]{AC10}. Generally speaking, it is used to evaluate how a statistical procedure, constructed on the basis of some training data, performs on a set of independent test data. Usually, the statistical procedure of interest depends on a number of tuning parameters, and hence cross-validation is typically employed to choose those tuning parameters in order to optimize the estimated performance.
	
	Here, we explore a different use of cross-validation, which apparently seems not to have been exploited before. In the following, this new approach is fully developed in the context of a specific problem in nonparametric estimation, though we note that its scope of applicability is much wider, and we also briefly examine other contexts where this principle yields new methodology.
	
	The base case on which we focus is the estimation of the integral of a squared density, since it represents the simplest problem to which our novel methodology can be applied.
	This functional arises when studying the efficiency of rank-based statistics, and in some other contexts like projection pursuit and symmetry testing or, more recently, in relation to causal estimation \citep{KBW20}. Moreover, minus one half of this quantity is known as differential extropy \citep{LSA15}, and can be considered as the complementary dual of differential entropy in the field of Information Theory \citep[see also][]{THB23}.
	
	This estimation problem has been studied by many authors, through different methodologies, and it is by now very well understood \citep[see][for a detailed account of contributions on the subject]{CT12}. Yet, it still continues to motivate new research, such as the investigations of \citet{GL22a,GL22b}.
	Among the existing estimators, a number of them are based on kernel smoothing; see \cite{SHD94}, and references therein, or \cite{GN08} and \cite{Wal14}, for more recent contributions. Other approaches make use of orthogonal series, as in \cite{L96}, \cite{K06} or \cite{T20}; or wavelets, as in \cite{KP96} and \cite{PR99}; or are motivated by model selection ideas, see \cite{LM00} and \cite{L05}. In addition, estimators based on spacings were studied in \cite{H84}, \cite{K89}, or \cite{vE92}.
	
	All those previous approaches share one feature in common: they depend on some tuning parameter that needs to be chosen (a bandwidth, a cutoff, a spacing order, etc). In contrast, the main novelty in our proposal is that it is fully empirical, in the sense that the new estimator is computed solely from the data, with no tuning parameters involved. Despite this simplicity, the method is shown to be efficient, in the sense that it reaches the information bound and the fastest convergence rate for the problem, provided the density is smooth enough.
	
	In Section 2 we develop the new methodology for the base problem, illustrate how the proposed estimator can be obtained via three different motivations, and accurately describe its large-sample behaviour. Section 3 contains a brief simulation study that shows the remarkable performance of the new proposal in practice. Section 4 deals with some selected extensions to highlight the wide applicability of the introduced principle. More intricate developments of the same idea are shown in Section 5 to yield novel, fully empirical selectors of the smoothing parameters for histogram and kernel density estimation. Finally, Section 6 contains a discussion on the implications of this new paradigm, and suggests further possibilities for future applications.

	\section{Estimation of the integral of a squared density}\label{sec:psiest}
	
	Consider independent and identically distributed univariate random variables $X_1,\dots,X_n$, with a common continuous distribution function $F$ with density $f$. The goal of this section is to present and study the properties of a new estimator of the statistical functional $\psi=\int_{\mathbb R} f(x)^2dx=\int_{\mathbb R}f(x)dF(x)$. As noted in the Introduction, this problem can be considered as the base case on which we will develop our novel methodology in full detail, but we will extend it in many other directions afterwards.
	
	%
	
	Following the terminology in \cite{JS91}, from the above integral expressions of $\psi$ two different but related `diagonals-in' kernel estimators can be proposed: first, $\widetilde \psi_{\rm D}^*(g)=\int_{\mathbb R} \widehat f(x;g)^2dx$, but also $\widetilde\psi_{\rm D}(g)=\int_{\mathbb R}\widehat f(x;g)dF_n(x)=n^{-1}\sum_{i=1}^n\widehat f(X_i;g)$. Here, $F_n$ denotes the empirical distribution function and $\widehat f(x;g)=n^{-1}\sum_{j=1}^nL_g(x-X_j)$ stands for the kernel density estimator, with kernel $L$ (a symmetric integrable function with unit integral) and bandwidth $g>0$, and $L_g(x)=L(x/g)/g$ is the scaled kernel. More explicitly,
	$$
	\widetilde \psi_{\rm D}^*(g)=n^{-2}\sum_{i,j=1}^n(L*L)_g(X_i-X_j)\quad\text{and}\quad
	\widetilde \psi_{\rm D}(g)=n^{-2}\sum_{i,j=1}^nL_g(X_i-X_j),
	$$
	where $*$ is the convolution product. This shows that both estimators are in fact of the same type, with the only difference that $\widetilde \psi_{\rm D}^*(g)$ employs the convolved kernel $L*L$ while $\widetilde \psi_{\rm D}(g)$ simply makes use of $L$. In addition, previously \cite{HM87} noted that $\widetilde\psi_{\rm D}(g)$ contains the non-stochastic term $n^{-1}L_g(0)$ and suggested to remove it and consider instead the closely-related `no-diagonals' estimator
	$\widetilde\psi_{\rm ND}(g)=\{n(n-1)\}^{-1}\sum_{i\neq j}L_g(X_i-X_j).$ Both estimators can be shown to be consistent if $g\equiv g_n$ satisfies $g\to0$ and $ng\to\infty$ as $n\to\infty$ \citep{BR69,GN08,CT12}.
	
	Nevertheless, the choice of $g$ in practice can be problematic: it is known that the optimal bandwidth depends on the integral of squared density derivatives of higher order \citep[][Section 3.5]{WJ95}, so this raises the problem of estimating these higher order functionals.  The options here are to use another kernel estimation stage to estimate those functionals, with some pilot bandwidth (and face again some further bandwidth selection problem), or, more simply, to adjust the smoothing level in order to be optimal for some location-scale family based on a reference distribution.
	In fact, in order to avoid an infinitely cyclic reasoning, the reference distribution approach must be employed at some initial step to start this process of multiple kernel stages. However, as noted in \cite{JMS96}, if a simple unimodal distribution is taken as a reference, then any bandwidth selector relying on such an initial stage will not pass a `bimodality test', in the sense that its performance could be made arbitrarily bad by sampling from a bimodal distribution with modes that are sufficiently far apart.
	
	Here we present a new estimator of $\psi$ which, despite being closely connected with the former kernel methods, does not require the specification of a bandwidth or any other tuning parameter. Thus, it allows us to overcome the bandwidth selection difficulties.
	
	The rationale for our novel proposal is the following: recall that to measure the performance of $\widehat f(\cdot;g)$ as an estimator of $f$ it is common to use the mean integrated squared error ${\rm MISE}(g)=\mathbb E\{{\rm ISE}(g)\}$, where ${\rm ISE}(g)=\int_{\mathbb R}\{\widehat f(x;g)-f(x)\}^2dx$. An unbiased estimator of ${\rm MISE}(g)-\psi$ is given by the cross-validation criterion \citep{R82,B84}, which can be explicitly written as
	$${\rm CV}(g)=(ng)^{-1}R(L)+\{n(n-1)\}^{-1}\sum_{i\neq j}\{(1-n^{-1})(L*L)_g-2L_g\}(X_i-X_j),$$
	where $R(\alpha)=\int_{\mathbb R}\alpha(x)^2dx$ for any square integrable function $\alpha$. Since ${\rm CV}(g)$ is unbiased for ${\rm MISE}(g)-\psi$ for each $g>0$ and $\min_{g>0}{\rm MISE}(g)\to0$ as $n\to\infty$ \citep{CMNP07}, it is reasonable to expect that $\min_{g>0}{\rm CV}(g)\to-\psi$ in probability. This suggests considering the estimator of $\psi$ defined as
	$$\widehat\psi=-\min_{g>0}{\rm CV}(g).$$
	Such a minimum is actually attained if $R(L)<2L(0)$ \citep{S84}, so all the kernels considered henceforth are assumed to satisfy this condition.
	
	A few remarks about this new estimator are in order: first, note that $\widehat\psi$ is fully empirical, in the sense that it does not require the specification of any further tuning parameter. Second, observe that $\widehat\psi$ makes quite an unconventional use of cross-validation: whereas its classical application focuses on its minimizer $\widehat g_{\rm CV}=\mathop{\rm argmin}_{g>0}{\rm CV}(g)$, which is seen as a data-based bandwidth that seeks to mimic the behaviour of $g_{\rm MISE}=\mathop{\rm argmin}_{g>0}{\rm MISE}(g)$, here the main interest lies in the minimum value that the cross-validation criterion attains, as an estimator of $-\psi$.
	
	Moreover, an additional motivation for $\widehat \psi$ can be given. Notice that the cross-validation criterion can be written in terms of the previously introduced kernel estimators of $\psi$, namely ${\rm CV}(g)=\widetilde\psi_{\rm D}^*(g)-2\widetilde\psi_{\rm ND}(g)$, so that
	\begin{equation}\label{eq:diff}
		\widehat\psi=-\min_{g>0}\{\widetilde\psi_{\rm D}^*(g)-2\widetilde\psi_{\rm ND}(g)\}=\max_{g>0}\{2\widetilde\psi_{\rm ND}(g)-\widetilde\psi_{\rm D}^*(g)\}.
	\end{equation}
	Since $\widetilde\psi_{\rm D}^*(g)$ and $\widetilde\psi_{\rm ND}(g)$ are consistent estimators of $\psi$ as long as $g\to0$ and $ng\to\infty$, then the same applies to the estimator $\widecheck\psi(g)=2\widetilde\psi_{\rm ND}(g)-\widetilde\psi_{\rm D}^*(g)$. But $\widehat g_{\rm CV}/g_{\rm MISE}\to1$ in probability \citep{H83}, with $g_{\rm MISE}\to0$ and $ng_{\rm MISE} \to\infty$ under fairly general assumptions \citep{CMNP07}, which means that we can expect $\widehat\psi=-{\rm CV}(\widehat g_{\rm CV})$ to be a consistent estimator of $\psi$.
	
	A third motivation for $\widehat\psi$ stems from the fact that we can also express
	\begin{equation}\label{eq:pen}
		\widehat\psi=\max_{g>0}\big\{\widetilde\psi_{\rm ND}^{\bullet}(g)-W(g)\big\},
	\end{equation}
	where $\widetilde\psi_{\rm ND}^\bullet(g)=\{n(n-1)\}^{-1}\sum_{i\neq j}(2L-L*L)_g(X_i-X_j)$ is of the same type as $\widetilde\psi_{\rm ND}(g)$, but based on the ``twicing'' kernel $2L-L*L$. Here, $W(g)=\widetilde\psi_{\rm ND}^\bullet(g)-2\widetilde\psi_{\rm ND}(g)+\widetilde\psi_{\rm D}^*(g)=(ng)^{-1}R(L)+O_P(n^{-1})$ is such that $0\leq W(g)\leq(ng)^{-1}R(L)$ if the kernel $L$ is positive \citep[][Lemma 3]{D89}, so it can be understood as a penalization term, and this unveils $\widehat\psi$ as closely connected to the estimators based on model selection \citep[see][]{LM00,L05}.

	Thus, $\widehat\psi$ is indeed based on kernel smoothing, but its most remarkable and striking feature is that its bandwidth is implicitly chosen, no multistage process or reference distribution are required. It remains intriguing, however, why the maximum of $\widecheck\psi(g)$ over the possible bandwidths should result in a good choice. The key to understand this feature lies in the behaviour of $\widecheck\psi(g)$ as an estimator of $\psi$, which is analyzed in detail in Appendix A of the Supplementary Material. The crucial observation is that minimizing the mean squared error of $\widecheck\psi(g)$ is asymptotically equivalent to minimizing its squared bias; this is an attribute that it shares with estimators of type $\widetilde\psi_{\rm D}(g)$ \citep{JS91,CT12}. Moreover, it is clear that the bias of $\widecheck\psi(g)=-{\rm CV}(g)$ equals $-{\rm MISE}(g)<0$; hence the bandwidth $g$ that minimizes the squared bias is precisely $g_{\rm MISE}$, which is in turn estimated by the cross validation bandwidth $\widehat g_{\rm CV}$ that is implicitly used in $\widehat \psi$.
	
	
	However, it is a common conclusion in comparative bandwidth selection studies that the cross-validation bandwidth is unacceptably unstable for its large variability \citep[see, e.g.,][]{CCGM94}, therefore, the fact that $\widehat\psi$ implicitly depends on $\widehat g_{\rm CV}$ may suggest that this could be the case for this estimator too. Notwithstanding, the next result shows another unforeseen and remarkable property of this estimator: it is consistent under minimal assumptions and, for a smooth enough density, it is asymptotically efficient since it is $\sqrt n$-consistent and attains the information bound for this problem.
	
	The smoothness of $f$ will be defined in terms of Sobolev spaces. Denote by $\varphi_f(t)=\int_\mathbb R e^{itx}f(x)dx$ the characteristic function associated with $f$ and define the Sobolev class of densities of order $\beta>0$ as $\mathcal H_\beta=\{\text{densities }f\text{ such that }\int_{\mathbb R}|t|^{2\beta}|\varphi_f(t)|^2dt<\infty\}$. The asymptotic behaviour of the new estimator $\widehat{\psi}$ is described in the next result.
	
	\begin{theorem}\label{thm:psi}
		If the density $f$ is bounded and $L$ is a symmetric density of bounded variation, continuous at zero, and such that $R(L)<2L(0)$ , then
		$$\Big(\widehat\psi-\psi-2n^{-1}\sum_{i=1}^nY_i\Big)\Big/{\rm MISE}(g_{\rm MISE})\to-1\quad\text{almost surely},$$
		where $Y_i=f(X_i)-\psi$. As a consequence, $\widehat\psi$ is a strongly consistent estimator of $\psi$.
		
		In addition, assume that $\int_{\mathbb R}x^2L(x)dx<\infty$ and $f\in\mathcal H_\beta$ for some $\beta>0$. If $\beta>1/2$, then
		$\sqrt{n}(\widehat\psi-\psi)\xrightarrow{\ d\ }N(0,4\tau^2),$
		where $\tau^2=\Var\{f(X_1)\}=\int_{\mathbb R}f(x)^3dx-\psi^2$. If $0<\beta\leq1/2$, then $\widehat\psi-\psi=O_P(n^{-2\beta/(2\beta+1)})$.
	\end{theorem}
	
	\begin{remark}\label{rem1}
		The proof of the first part of Theorem \ref{thm:psi} is given in Section \ref{sec:proofs} below. It follows that $\widehat\psi-\psi=2n^{-1}\sum_{i=1}^nY_i+O_P\{{\rm MISE}(g_{\rm MISE})\}$ so, asymptotically, $\widehat\psi-\psi$ behaves as (twice) an average of centred random variables plus a remainder. To prove the second statement of the theorem, note that the first term is easily described with a central limit theorem, so it suffices to study the rate of convergence of ${\rm MISE}(g_{\rm MISE})$. Proceeding as in the proof of Theorem 1.5 in \cite{T09} it can be shown that ${\rm MISE}(g_{\rm MISE})=O(n^{-2m/(2m+1)})$, where $m=\min\{\beta,2\}$. Hence, the remainder is indeed negligible if $\beta>1/2$, but otherwise becomes the dominant term, in which case it is of order $O(n^{-2\beta/(2\beta+1)})$.
	\end{remark}
	
	\begin{remark}
		Unfortunately, Theorem \ref{thm:psi} shows that the estimator $\psi$ is not rate-adaptive. \cite{BR88} showed that $\psi$ can be estimated at rate $n^{-1/2}$ if $f$ is $\beta$-H\"{o}lder with $\beta>1/4$ and at rate $n^{-4\beta/(4\beta+1)}$ if $\beta\leq1/4$; however, in the related Sobolev framework, the estimator $\widehat\psi$ needs $\beta>1/2$ to be asymptotically efficient. In any case, this is not a too restrictive smoothness assumption since it holds if $f$ is absolutely continuous and its derivative is square integrable. Furthermore, $\widehat\psi$ is still $\sqrt n$-consistent (though with an asymptotic variance greater than $4\tau^2$) if $\beta=1/2$, which covers the case of a density with a finite number of discontinuity points \citep{vE85}. Again, the main advantage of $\widehat\psi$ is that it does not require any tuning parameter to be chosen.
	\end{remark}
	
	\begin{remark}
		As noted above, the motivation of $\widehat \psi$ in terms of a penalized criterion reveals some interesting connections with the estimation of $\psi$ based on model selection. In that framework, a judicious choice of the penalization allows for the construction of rate-adaptive estimators; therefore, this raises the open question of whether a suitable modification of the penalty term in (\ref{eq:pen}) could yield a rate-adaptive estimator in this context as well.
	\end{remark}
	
	
	\section{A brief numerical study}
	
	To explore the performance of the new estimator $\widehat\psi$ in practice, a brief simulation study was carried out. Due to their nice convolution properties, the test densities included in this study were the 15 normal mixture densities introduced in \cite{MW92}, plus the 10-modal normal mixture described in \cite{L99}. 
	
	For the sake of brevity, only the behaviour of three estimators of $\psi$ is reported here: the 2-stage solve-the-equation kernel estimator of \cite{SHD94}, the 2-stage direct-plug-in kernel estimator of \cite{JS91} and the new method introduced in Section \ref{sec:psiest}. They are labeled as $\widehat\psi_{\rm SHD}$, $\widehat\psi_{\rm JS}$ and $\widehat\psi_{\rm CT}$, respectively, in the sequel. In all cases, the standard normal density was used as the kernel. These three estimators share the feature that the choice of all the tuning parameters required for their implementation (none, in the case of the new proposal) were clearly specified in their respective references.
	
	Another estimator that fulfils that condition is the one proposed in \citet{W95}; however, it showed an unsatisfactory performance in some preliminary experiments and was left out of the final comparison. Many other proposals in the literature, even if shown to enjoy nice theoretical properties, depended on some unspecified constants, with no explicit indication on how to choose them in practice, and for this reason they were not included in the study either.
	
	Two sample sizes were explored, $n=100$ and $n=1000$. For each of these two scenarios, $B=500$ samples of size $n$ from each test density were drawn. The performance of a certain estimator of $\psi$, say $\widehat\psi_{\rm XY}$, with regard to some test density $f$, was measured by the relative root mean squared error, ${\rm RRMSE}=\{B^{-1}\sum_{b=1}^B(\widehat\psi_{{\rm XY},b}-\psi)^2\}^{1/2}/\psi$, where $\widehat\psi_{{\rm XY},b}$ is the estimate obtained from the $b$-th synthetic sample.
	
	To compare the performance of the estimators in the study, for each test density we computed the RRMSE of each estimator divided by the minimum RRMSE among the three of them for that density model. Such a ratio represents how bad an estimator is with respect to the best, which corresponds to a ratio value of $1.000$. An extended description with more detailed results for each density model and sample size is included in Appendix B of the Supplementary Material. Here, in Table \ref{tab:rrmse}, to save space we only include four summary statistics for the 16 test densities: the mean, median, minimum and maximum RRMSE ratio for each estimator.
	
	\begin{table}[!ht]
		\centering
		\begin{tabular}{lccc|ccc}
			&\multicolumn{3}{c}{$n=100$}&\multicolumn{3}{c}{$n=1000$}\\\cline{2-7}
			&CT&SHD&JS&CT&SHD&JS\\\hline\hline
			Mean    & 1.095 & 1.261 & 1.335 & 1.044 & 1.300 & 2.246 \\
			Median  & 1.080 & 1.113 & 1.007 & 1.019 & 1.029 & 1.032 \\
			Minimum & 1.000 & 1.000 & 1.000 & 1.000 & 1.000 & 1.000 \\
			Maximum & 1.262 & 3.778 & 3.828 & 1.220 & 5.219 & 16.672
		\end{tabular}
		\caption{Summary statistics for the RRMSE ratios of the three compared methods, for the 16 test densities. A smaller ratio corresponds to a lower error.}
		\label{tab:rrmse}
	\end{table}
	
	The main conclusion from Table \ref{tab:rrmse} is that the new method is always the best or very close to the best in terms of performance. On average, it is about 10\% worst than the best for $n=100$ and 5\% worst than the best for $n=1000$. The fact that the minimum among the 16 density models is 1.000 for all the methods indicates that each of them is the best for at least one of the considered test densities.
	
	The relatively high maximum ratios observed for the estimators $\widehat\psi_{\rm SHD}$ and $\widehat\psi_{\rm JS}$ both correspond to the case of Loader's ten-modal density, for which $\widehat\psi_{\rm CT}$ performs best. While $\widehat\psi_{\rm CT}$ obtains reasonably good results already for $n=100$, the other two show a relatively bad performance even for $n=1000$. It is worth remarking that this is a density model for which the `local variability' in each of the ten components is very poorly modelled by a global, unimodal reference, and this may be the crucial factor hindering the performance of $\widehat\psi_{\rm SHD}$ and $\widehat\psi_{\rm JS}$ (which eventually rely on a normal-reference initial stage).

	\section{Extensions}\label{sec:ext}
	
	A number of extensions of the main new idea are possible, which shows its applicability in other contexts. In the following two sections we briefly explore some of these possibilities.
	
	\subsection{Directional data}
	
	The use of a reference distribution in an initial estimation step is especially problematic for the case of directional data. Attempting to fit circular data using a simple unimodal reference distribution, when they were truly drawn from a density with two antipodal modes, often leads to a fit that is close to the uniform distribution \citep[see][]{OCR12} and, therefore, to severely oversmoothed kernel estimates if they are based on the reference distribution approach.
	
	Next we show how our new methodology can be easily adapted to estimate the integral of the squared density of a circular distribution. Let $\Theta_1,\dots,\Theta_n$ be a random sample from a circular density $f\colon[0,2\pi)\to\mathbb R$. The goal here is to estimate $\psi=\int_0^{2\pi}f(\theta)^2d\theta$, and we will rely on the kernel density estimator in this context, given by $\widehat f(\theta;\nu)=n^{-1}\sum_{i=1}^nK_\nu(\theta-\Theta_i)$, where now $\nu>0$ is a concentration parameter, which acts as the bandwidth, and $K_\nu$ is a circular kernel function. Quite commonly, the von Mises kernel with concentration $\nu$ is employed, so that $K_\nu(\theta-\Theta_i)=\exp\{\nu(\theta-\Theta_i)\}/\{2\pi I_0(\nu)\})$, where $I_0$ denotes the 0-order modified Bessel function of the first kind  \citep[see][]{T08}.
	
	The cross-validation criterion ${\rm CV}(\nu)$ for density estimation with spherical data was introduced in \cite{HWC87}. There, it was shown that its minimum expected value satisfies $\min_{\nu>0} \mathbb E\{{\rm CV}(\nu)\}\to -\psi$ as $n\to\infty$. Thus, arguing as in the linear case above, a reasonable estimator of $\psi$ can be defined as $\widehat \psi=-\min_{\nu>0}{\rm CV}(\nu)$.
	
	The practical performance of this estimator was investigated in \citet{Ch17}, by comparing it with the usual kernel estimator proposed in \citet{DMPT11} on the basis of the 20 circular mixture models introduced in \citet{OCR12}. The new estimator performed best or second best (and quite close to the best) for all the models. For those distributions with antipodal modes, the classical 2-stage kernel estimator with an initial von Mises reference showed a specially bad performance, as expected due to the aforementioned fitting issue; meanwhile, on the contrary, those models posed no problem at all for the new proposal, which does not rely on any reference distribution.

	\subsection{Measurement errors}
	
	The problem of estimating $\psi$ has also been considered for the case in which the data are subject to measurement errors \citep{DG02}. In that context, instead of observing the error-free variables $X_1,\dots,X_n$, we are given a sample $Y_1,\dots,Y_n$ such that $Y_j=X_j+\varepsilon_j$, where the unobservable errors $\varepsilon_1,\dots,\varepsilon_n$ are independent of $X_1,\dots,X_n$ and are assumed to have a fully-known density $f_\varepsilon$. Hence, the problem is that we are interested in estimating $\psi=\int_{\mathbb R}f_X(x)^2dx$ but we observe a sample with density $f_Y=f_X*f_\varepsilon$, where $f\equiv f_X$ in this section.
	
	The classical approach here is based on deconvolution kernel density estimators: noting that the characteristic functions of $Y$, $X$ and $\varepsilon$ are related through $\varphi_Y=\varphi_X\varphi_\varepsilon$ with fully known $\varphi_\varepsilon$, then by Fourier inversion $f_X$ is estimated by $\widehat f_X(x;g)=(ng)^{-1}\sum_{i=1}^nK_\varepsilon(\frac{x-Y_i}g;g)$, where $K_\varepsilon(x;g)=(2\pi)^{-1}\int_{\mathbb R} \exp(-itx)\varphi_K(t)/\varphi_\varepsilon(t/g)dt$, with $\varphi_K(t)=\int_{\mathbb R}\exp(itx)K(x)dx$ standing for the Fourier transform of a kernel function $K$. The estimator of $\psi$ studied in \cite{DG02} is then $\widetilde \psi_D^*(g)=\int_\mathbb R\widehat f_X(x;g)^2dx$, by analogy with the error-free case. A multi-stage procedure with a parametric start is proposed by those authors to select the bandwidth $g$.
	
	In contrast, \cite{H99} derived a cross-validation criterion ${\rm CV}(g)$ for the density estimator $\widehat f_X(x;g)$ which, as in the error-free case, satisfies $\mathbb E[{\rm CV}(g)]={\rm MISE}(g)-\psi$, where now ${\rm MISE}(g)=\mathbb E\int_{\mathbb R}\{\widehat f_X(x;g)-f_X(x)\}^2dx$. Therefore, a bandwidth-free estimator can be analogously defined as $\widehat \psi=-\min_{g>0}{\rm CV}(g)$ in this context. In this case, both the theoretical and practical investigation of this estimator are open for future research.

	\subsection{Entropy estimation}

Entropy estimation is a topic of great interest, both for statisticians \citep[see][]{BSY19,HJWW20} and within the Information Theory community and other areas (see the survey of \citealp{V19}). A number of nonparametric entropy estimators and their applications were reviewed in \cite{Bal97}.

The entropy of a distribution with density $f$ is defined as $H=-\int_\mathbb R f(x)\log f(x)dx$. This statistical functional presents some similarities with $\psi$. For instance, it can be alternatively written as $H=-\int_{\mathbb R}\log f(x)dF(x)$,  and this suggests two possible estimators: first, $\widetilde H_{\rm D}^*(g)=-\int_\mathbb R \widehat f(x;g)\log \widehat f(x;g)dx$, but also $\widetilde H_{\rm D}(g)=-n^{-1}\sum_{i=1}^n\log\widehat f(X_i;g)$. The latter is more commonly used in practice, since it avoids numerical integration. Both of them contain some non-stochastic terms, so a third possibility is to use $\widetilde H_{\rm ND}(g)=-n^{-1}\sum_{i=1}^n\log\widehat f_i(X_i;g)$, where $\widehat f_i(x;g)=\{(n-1)h\}^{-1}\sum_{j\neq i}K\{(x-X_j)/h\}$.

Bandwidth selection for these estimators is a difficult and relatively unexplored topic. However, the methodology hereby introduced also circumvents this problem as follows: the last of the previous three estimators satisfies $\widetilde H_{\rm ND}(g)=-{\rm LCV}(g)$, where the latter acronym stands for likelihood cross-validation, a criterion related to measuring the density estimation error through the Kullback-Leibler divergence ${\rm KL}(g)=\int_\mathbb Rf(x)\log\{f(x)/\hat f(x;g)\}dx$. In fact, \cite{H87} showed that $\mathbb E[\widetilde H_{\rm ND}(g)]=-\mathbb E[{\rm LCV}(g)]=\mathbb E[{\rm KL}(g)]+H$ so, since $0\leq\min_{g>0}\mathbb E[{\rm KL}(g)]\to0$ under appropriate conditions, it follows that a reasonable, fully empirical entropy estimator is given by
$$\widehat H=\min_{g>0}\widetilde H_{\rm ND}(g)=-n^{-1}\max_{g>0}\sum_{i=1}^n\log\widehat f_i(X_i;g).$$
It must be noted that this coincides exactly with the bandwidth selection recommendation for $\widetilde H_{\rm ND}(g)$ given in \citet[][Section 3]{HM93}, albeit with a slightly different motivation. Thus, the theoretical properties and practical performance of $\widehat H$ can be consulted in that paper.

Recently, \cite{DG22} posed the open problem of finding a fully data-driven entropy estimate that is consistent under the only condition that $H<\infty$. Given that cross-validation techniques are typically consistent under minimal conditions, we believe that $\widehat H$ is a firm candidate to satisfy the requirements of that open problem.

\subsection{Estimation of integrated squared density derivatives}

The problem of estimating $\psi$ is a particular instance of the more general problem of estimating the integrated squared $r$-th density derivative $\theta_r=\int_{\mathbb R}f^{(r)}(x)^2dx$. This functional is of great interest for the problem of automatic smoothing parameter selection for density estimation, especially the cases $r=1$ for the histogram estimator, and $r=2$ for the kernel estimator \citep[see][]{HM87,JS91,W97}.

This more general problem is connected to the estimation of the $r$-th density derivative $f^{(r)}(x)$. The usual estimator is constructed by taking the $r$-th derivative of the kernel density estimator $\widehat f^{(r)}(x;g)=(ng^{r+1})^{-1}\sum_{i=1}^nL^{(r)}\{(x-X_i)/g\}$, and its error is globally measured through ${\rm MISE}_r(g)=\mathbb E\{{\rm ISE}_r(g)\}$, where ${\rm ISE}_r(g)=\int_\mathbb R\{\widehat f^{(r)}(x;g)-f^{(r)}(x)\}^2dx$. For kernel density derivative estimation, the cross-validation criterion ${\rm CV}_r(g)$ was first introduced by \cite{HMW90}. It is shown in \cite{CD13} that $\mathbb E[{\rm CV}_r(g)]={\rm MISE}_r(g)-\theta_r$ so, since $\min_{g>0}{\rm MISE}_r(g)\to0$ \citep{CDW11}, again this justifies defining the fully empirical estimator of $\theta_r$ as $\widehat \theta_r=-\min_{g>0}{\rm CV}_r(g)$.

The theoretical properties of $\widehat \theta_r$ could be deduced by following the same steps as for the case $r=0$ (i.e., the estimation of $\psi$) in Theorem \ref{thm:psi} and Remark \ref{rem1}. The error $\widehat\theta_r-\theta_r$ can be written as an average of centred random variables plus a remainder of the same order as $\min_{g>0}{\rm MISE}_r(g)$. The latter can be shown to be $O(n^{-2m/(2m+2r+1)})$ when $f^{(r)}\in\mathcal H_\beta$ with $\beta>r$ and $L$ has a finite second-order moment, where $m=\min\{\beta-r,2\}$, thus leading to a $\sqrt n$-consistent estimator whenever $\beta\geq 2r+1/2$, provided $r\in\{0,1\}$. However, preliminary numerical work seems to suggest that the performance of $\widehat\theta_r$ might not be as satisfactory for $r\geq1$ as for $r=0$, so this remains an avenue for further research.

\section{Applications to density estimation}\label{sec:de}

All the previous extensions deal with the problem of estimating a real-valued statistical functional. In a step further, in terms of complexity, this section investigates how to apply the new methodology for the estimation of the error function of density estimators.

\subsection{Bandwidth selection for kernel density estimation}\label{sec:bw_kde}

Here we show how an application of the same principles introduced in this paper could also  be useful to suggest a new bandwidth selector for the classical kernel density estimator, closely related to the smoothed cross validation methodology, but with an automatic, implicit choice of the pilot bandwidth.

Consider now the kernel density estimator $\widehat f(x;h)=n^{-1}\sum_{i=1}^nK_h(x-X_i)$ with kernel $K$ and bandwidth $h>0$. As before, let us measure its performance in terms of ${\rm MISE}(h)=\mathbb E\int_{\mathbb R}\{\widehat f(x;h)-f(x)\}^2dx$ and denote the optimal level of smoothing in this sense by $h_{\rm MISE}=\mathop{\rm argmin}_{h>0}{\rm MISE}(h)$. The goal is to find an estimator of the MISE function, which would suggest an automatic bandwidth choice by minimizing such an error estimate.

First, notice that  the MISE can be expressed as
\begin{equation}\label{eq:exactMISE}
	{\rm MISE}(h)=\{(nh)^{-1}R(K)-n^{-1}R(K_h*f)\}+\{R(K_h*f)-2R_{K,h}(f)+R(f)\},
\end{equation}
where we are denoting $R(\alpha)=\int_{\mathbb R}\alpha(x)^2dx$ for any square integrable function $\alpha$ and $R_{\alpha,h}(f)=\int_{\mathbb R}(\alpha_h*f)(x)f(x)dx$ for any integrable function $\alpha$ \citep{CD18}. The first term in curly brackets in (\ref{eq:exactMISE}) corresponds to the integrated variance of $\widehat f(\cdot;h)$, while the second one is the integrated squared bias.

The MISE is usually simplified by keeping only the dominant term of the integrated variance, namely $(nh)^{-1}R(K)$, since this can be shown to have no effect on bandwidth selection \citep[][Theorem 1]{CD11}. This leads to the criterion
\begin{equation}\label{eq:M}
	M(h)=(nh)^{-1}R(K)+\{R(K_h*f)-2R_{K,h}(f)+R(f)\},
\end{equation}
so that the problem of estimating the MISE reduces to estimating the integrated squared bias. In addition, note that $R(K_h*f)=R_{K*K,h}(f)$ for symmetric $K$. Therefore, it suffices to investigate how to estimate functionals of the type $R_{\alpha,h}(f)$.



It is convenient to introduce here the notation $K_0$ for the Dirac delta function, as in \cite{JMP91}. It is not a proper function, but a generalized function \citep[see][]{GS64} which acts within an integral so that $\int_{\mathbb R}\alpha(x)K_0(x)dx=\alpha(0)$ for any integrable function $\alpha$; in particular, this implies that $K_0*\alpha=\alpha$. The notation $K_0$ is well suited because $K_h*\alpha(x)\to \alpha(x)=K_0*\alpha(x)$ as $h\to0$ if $x$ is a point of continuity of a bounded function $\alpha$ \citep[][Theorem 9.8]{WZ77}. With this notation, the error criterion (\ref{eq:M}) can be simply written as
\begin{equation}\label{eq:M2}
	M(h)=(nh)^{-1}R(K)+R_{K*K-2K+K_0,h}(f),
\end{equation}
so  it suffices to estimate $R_{\alpha,h}(f)$ for this particular instance of $\alpha=K*K-2K+K_0$.

Next, by writing $R_{\alpha,h}(f)$ as $\psi_{\alpha,h}=\int_\mathbb R(\alpha_h*f)(x)dF(x)$ it is immediate to recognize its similarity with $\psi=\int_\mathbb Rf(x)dF(x)$. The latter is the same as the former, except for the convolution with $\alpha_h$ in the integrand and, besides, for the Dirac delta $\alpha_h=K_0$ we have $\psi_{\alpha,0}=\psi$. Therefore,  the previously introduced methodology regarding $\psi$ can be adapted to estimate $\psi_{\alpha,h}$ and, hence, the error criterion $M(h)$.

For instance, the smoothed cross validation criterion \citep{HMP92} is obtained by estimating $R_{\alpha,h}(f)=\psi_{\alpha,h}$ in (\ref{eq:M2}) by
\begin{equation}\label{eq:psi.alpha}
	{\widetilde\psi}_{\alpha,h}^*(g)=\int_\mathbb R\big\{\alpha_h*\widehat f(\cdot;g)\big\}(x)\widehat f(x;g)dx=n^{-2}\sum_{i,j=1}^n\{\alpha_h*(L*L)_g\}(X_i-X_j),
\end{equation}
thus leading to ${\rm SCV}(h;g)=(nh)^{-1}R(K)+{\widetilde\psi}_{K*K-2K+K_0,h}^*(g)$ as an estimator of $M(h)$. A very careful choice of the pilot bandwidth $g$, depending on $h$, can make the SCV selector $\widehat h_{{\rm SCV}}\equiv\widehat h_{{\rm SCV}}(g)=\mathop{\rm argmin}_{h>0}{\rm SCV}(h;g)$ attain the fastest possible relative convergence rate towards $h_{\rm MISE}$ \citep{JMP91}.

The choice of such a pilot bandwidth $g$, though, is a very delicate issue that involves a multi-stage kernel estimation process and, eventually, the use of an initial reference distribution, with all the aforementioned problems associated to this strategy. However, with the new methodology hereby introduced, it is possible to define an estimator of $\psi_{\alpha,h}$ which makes use of an implicitly defined pilot smoothing level, so that it does not need the specification of any further tuning parameters (although it is also based on kernel smoothing). Let us elaborate on this.

Any of the three motivations for the estimator $\widehat \psi$ of $\psi$ can be adapted for the estimation of $\psi_{\alpha,h}$, but perhaps the most straightforward technique is the one that combines the diagonals-in estimator introduced in (\ref{eq:psi.alpha}) with the no-diagonals estimator. Explicitly, the new, bandwidth-free estimator of $\psi_{\alpha,h}$ is defined as
$${\widehat\psi}_{\alpha,h}=\max_{g>0}\Big\{\frac{2}{n(n-1)}\sum_{i\neq j}(\alpha_h*L_g)(X_i-X_j)-\frac{1}{n^{2}}\sum_{i,j=1}^n\{\alpha_h*(L*L)_g\}(X_i-X_j)\Big\}.$$
By using this estimator we obtain a fully empirical estimator of the error criterion $M(h)$, namely $\widehat M(h)=(nh)^{-1}R(K)+\widehat \psi_{K*K-2K+K_0,h}$. The corresponding novel bandwidth selector $\widehat h=\mathop{\rm argmin}_{h>0}\widehat M(h)$ can thus be seen as a variant of the SCV approach, but with the remarkable difference that no pilot bandwidth choice is needed for its implementation.

Not relying on any auxiliary tuning parameter, the new $\widehat h$ can be easily implemented (once the involved double optimization process is handled carefully), and its practical performance can be inspected in a simulation study. Indeed, some preliminary results were reported in \citet{Ch15}, showing that $\widehat h$ effectively reduces the instability of the classical cross-validation selector for simple models (as expected from its pilot pre-smoothing feature), while at the same time it does not exhibit a too large bias in more complex scenarios (since it does not need an initial stage depending on a reference distribution). Its theoretical analysis, however, is much more complicated than usual (partly because of the double optimization as well), so for the moment it stands as an open problem.

\subsection{Histogram estimates}

Another popular density estimator is the histogram. Even if it is by now well-understood that histograms are not as efficient as kernel density estimators, they still stand among the most commonly used density estimators, due to their simplicity. A detailed study of histograms as density estimators is contained in Chapter 3 of \citet{S15}.

To construct a histogram the first step is to divide the real line into bins $\{B_k\}_{k\in\mathbb Z}$. To fix ideas we will simply consider the anchor point to be the origin, and all bins to have the same binwidth $b$, so that $B_k=[kb,(k+1)b)$. Then, the histogram density estimator is defined as $\widetilde f(x;b)=(nb)^{-1}\sum_{k\in\mathbb Z}\nu_kI_{B_k}(x)$, where $I_A$ denotes the indicator function of a set $A$ and $\nu_k\equiv\nu_k(b)$ is the number of sample points falling within $B_k$.

The MISE of the histogram estimate can be exactly written as
\begin{equation}\label{eq:MISEhist}
	{\rm MISE}(b)=(nb)^{-1}-(nb)^{-1}(n+1)\sum_{k\in\mathbb Z} p_k^2+R(f),
\end{equation}
where $p_k\equiv p_k(b)=\int_{B_k}f(x)dx$. On the other hand, its shifted version ${\rm MISE}(b)-R(f)$ is unbiasedly estimated by ${\rm CV}(b)=2/\{(n-1)b\}-(n+1)/\{n^2(n-1)b\}\sum_{k\in\mathbb Z}\nu_k^2$ \citep{S15}. So again, as in Section \ref{sec:psiest}, a sensible, fully empirical estimate of $R(f)$ can be proposed as $\breve\psi=-\min_{b>0}{\rm CV}(b)$.

To highlight the similarities of the histogram-type estimate $\breve\psi$ and the kernel-based estimate $\widehat\psi$, it is convenient to note that
$$\sum_{k\in\mathbb Z}\nu_k^2=\sum_{i,j=1}^n\Big\{\sum_{k\in\mathbb Z}I_{B_k}(X_i)I_{B_k}(X_j)\Big\}=\sum_{i\neq j}\Big\{\sum_{k\in\mathbb Z}I_{B_k}(X_i)I_{B_k}(X_j)\Big\}+n.$$
The former shows how the $V$-statistic $V(b)=(n^2b)^{-1}\sum_{k\in\mathbb Z}\nu_k^2$ can be expressed as $U(b)+(nb)^{-1}$, where $U(b)=(n^2b)^{-1}\sum_{i\neq j}\Big\{\sum_{k\in\mathbb Z}I_{B_k}(X_i)I_{B_k}(X_j)\Big\}$ is the corresponding $U$-statistic. Hence, modulo $n-1\approx n\approx n+1$ we can write
$$\breve\psi\approx\max_{b>0}\big\{V(b)-2/(nb)\big\}=\max_{b>0}\big\{2U(b)-V(b)\big\},$$
which is the equivalent to (\ref{eq:diff}) for the histogram estimate.

The cross-validation criterion here is derived by replacing the unknown term $b^{-1}\sum_{k\in\mathbb Z}p_k^2$ in the MISE expression with the empirical estimate $U(b)$. Following the analogy with the kernel estimator, a smooth cross-validation criterion is obtained when $b^{-1}\sum_{k\in\mathbb Z}p_k^2$ is estimated thorugh $V_b(c)=b^{-1}\sum_{k\in\mathbb Z}\{\int_{B_k}\widetilde f(x;c)dx\}^2$, where $\widetilde f(x;c)$ is the histogram density estimator with pilot binwidth $c$. More explicitly, if $\widetilde f(x;c)$ is based on the partition $\{C_\ell\}_{\ell\in\mathbb Z}$ with $C_\ell=[\ell c,(\ell+1)c)$, then
$\int_{B_k}\widetilde f(x;c)dx=(nc)^{-1}\sum_{i=1}^n\sum_{\ell\in\mathbb Z}I_{C_\ell}(X_i)\lambda(B_k\cap C_\ell)$ where $\lambda$ denotes the Lebesgue measure in $\mathbb R$; therefore,
\begin{equation}\label{eq:Vbc}
	V_b(c)=b^{-1}(nc)^{-2}\sum_{i,j=1}^n\sum_{k\in\mathbb Z}\sum_{\ell,\ell'\in\mathbb Z}I_{C_\ell}(X_i)I_{C_{\ell'}}(X_j)\lambda(B_k\cap C_\ell)\lambda(B_k\cap C_{\ell'}).
\end{equation}

The main difficulty here lies on the choice of the pilot binwidth $c$. However, constructing the corresponding $U$-statistic $U_b(c)$ as in (\ref{eq:Vbc}), but with the first double sum restricted to $i\neq j$, and reasoning analogously as for the kernel estimator, a sensible estimate of the unknown $b^{-1}\sum_{k\in\mathbb Z}p_k^2$ is given by $\max_{c>0}\{2U_b(c)-V_b(c)\}$. Replacing this estimate in (\ref{eq:MISEhist}) gives a smooth cross validation criterion for the histogram, which automatically (internally) adjusts the pilot binwidth, and hence leads to a new, fully empirical, SCV-type binwidth selector for histogram density estimation.

\section{Discussion}

When a certain methodology depends on some tuning parameter, and cross-validation is employed to estimate its performance, its most common utility lies in using its minimiser as a data-driven choice for the tuning parameter. Here, a different feature is explored, namely how the corresponding optimal cross-validated performance defines an estimator of a certain statistical functional.

This approach is fully developed for the base case of cross-validation for kernel density estimation, where a new estimator of the integrated squared density is found, with the remarkable property that it does not rely on the choice of any further tuning parameter. The whole analysis is conducted in the univariate setting, but it could be extended in a straightforward manner to the multivariate case, where the standard kernel estimate of this functional does depend on the choice of a smoothing parameter \citep[see][Section 3.2]{CD10}.

In addition, here the focus is on independent data, but it can be noted that cross-validation techniques for density estimation with dependent data are also available \citep[see][]{HV90}, so an analogous procedure for the estimation of the integrated squared density with dependent data could be derived in a similar fashion.

Other applications are outlined in Sections \ref{sec:ext} and \ref{sec:de}, which shows the potential of the novel approach. Those related to smoothing parameter selection for density estimation surely deserve further attention, from both theoretical and practical perspectives. Particularly, it appears that the bandwidth selector introduced in Section \ref{sec:bw_kde} might have some connection to the methodology developed by \cite{GL11}, albeit with a completely different motivation. Studying its asymptotic and finite-sample properties constitutes a future research challenge.

\section{Proof of Theorem \ref{thm:psi}}\label{sec:proofs}


The proof of Theorem \ref{thm:psi} makes extensive use of the results in \cite{NP87}, which show that several random criteria $C(g)$, say, are almost surely uniformly equivalent to ${\rm MISE}(g)$. Precisely, this means that $\sup_{g>0}|C(g)/{\rm MISE}(g)-1|\to0$ almost surely.

To begin with, under the conditions of Theorem \ref{thm:psi}, Equations (12) and (13) in \cite{NP87} together assure that
\begin{equation*}\label{eq:reldif}
	\sup_{g>0}\left|\frac{{\rm MISE}(g)-{\rm CV}(g)-\psi-Z_n}{{\rm MISE}(g)}\right|\to0\quad\text{ almost surely},
\end{equation*}
where $Z_n=2n^{-1}\sum_{i=1}^n Y_i$. Since ${\rm MISE}(g)>0$ for all $g>0$ then, arguing as in \citet[][p. 794]{NP87}, we have
\begin{align*}
	\{1-o(1)\}{\rm MISE}(\widehat g_{\rm CV})&={\rm CV}(\widehat g_{\rm CV})+\psi+Z_n\\
	&\leq{\rm CV}(g_{\rm MISE})+\psi+Z_n=\{1+o(1)\}
	{\rm MISE}(g_{\rm MISE}).
\end{align*}
But also ${\rm MISE}(g_{\rm MISE})\leq {\rm MISE}(\widehat g_{\rm CV})$, so it follows that 
$$1-o(1)\leq\frac{{\rm CV}(\widehat g_{\rm CV})+\psi+Z_n}{{\rm MISE}(g_{\rm MISE})}\leq1+o(1),$$
and we can conclude that
$$\frac{{\rm CV}(\widehat g_{\rm CV})+\psi+Z_n}{{\rm MISE}(g_{\rm MISE})}\to1\quad\text{ almost surely},$$
which finishes the proof of Theorem \ref{thm:psi}.

\bigskip

\noindent{\bf Acknowledgement.} We are grateful to an associate editor and a reviewer for their remarks, which led to sensible presentation improvements.

\bigskip

\noindent{\bf Funding.} J. E. Chac\'{o}n has been partially
supported by Spanish Ministerio de Ciencia e Innovaci\'on grants PID2019-109387GB-I00 and PID2021-124051NB-I00, C. Tenreiro has been partially supported by the Centre for Mathematics of the University of Coimbra (funded by the Portuguese Government through FCT/MCTES, doi: 10.54499/UIDB/00324/2020).

\bigskip

\noindent{\bf Code.} An R script implementing the estimators introduced in Sections \ref{sec:psiest} and \ref{sec:bw_kde} is available at the webpage of the first author ({\tt matematicas.unex.es/\~{}jechacon}).

\bibliographystyle{apalike}

\newpage 

\setcounter{page}{1}

\begin{center}
	\Large Supplementary material to\\ ``Estimation of density functionals via cross-validation''
\end{center}


\begin{center}

\large

\author{
	Jos\'e E. Chac\'on\footnote{IMUEx, Departamento de
		Matem\'aticas, Universidad de Extremadura, Badajoz, Spain. E-mail:
		{\tt jechacon@unex.es}} \ and Carlos Tenreiro\footnote{CMUC, Department of
		Mathematics, University of Coimbra, Coimbra, Portugal. E-mail: {\tt tenreiro@mat.uc.pt}}
}

\end{center}

	
	\begin{abstract}
		\noindent  This document contains two appendices as supplementary material to the manuscript
		{\it  Estimation of density functionals via cross-validation}. Appendix \ref{app:a} provides a detailed theoretical study of one of the auxiliary estimators of the integrated squared density considered in the paper. Appendix \ref{app:b} supplies an elaborated analysis of the results of the simulation study contained in the paper.

	\end{abstract}
	
	\newpage
	
	\appendix
	
	\section{Detailed theoretical analysis of $\check\psi(g)$}\label{app:a}
	
	In this appendix we provide a detailed analysis of the estimator of
	$\psi=\int_{\mathbb R} f(x)^2 dx$ defined by
	$\widecheck\psi(g)=-{\rm CV}(g)$, where ${\rm CV}$ is the cross-validation function
	associated to the kernel density estimator $\widehat f(\cdot;g)$ of $f$. More precisely,
	we prove that there exists a bandwidth $g_{\rm MSE}$ that minimizes the mean squared error ${\rm MSE}(g)$ of $\widecheck\psi(g)$
	and that it is asymptotically equivalent to the bandwidth $g_{\rm MISE}$ that minimizes the mean integrated squared error of $\widehat f(\cdot;g)$.
	These results support the idea of taking $g_{\rm MISE}$ as the target bandwidth for $\widecheck\psi(g)$,
	giving us an additional motivation for estimating $\psi$
	by $\widehat\psi = \widecheck\psi(\widehat{g}_{\rm CV})$, where the cross-validation bandwidth $\widehat{g}_{\rm CV}=\mathop{\rm argmin}_{g>0}{\rm CV}(g)$ seeks to mimic the behaviour of $g_{\rm MISE}$.
	
	\subsection{Introduction}
	
	Given $X_1,\dots,X_n$ independent real-valued random variables with common probability density function $f$,
	we are interested in the estimation of the functional
	$$\psi = \int_{\mathbb R} f(x)^2 dx.$$
	For that we consider the estimator of $\psi$ given by
	$$\widecheck\psi(g)=-{\rm CV}(g),$$
	where $g\equiv g_n>0$ is a sequence of real numbers converging to zero as $n$ tends to infinity, and
	${\rm CV}(g)$, given by
	$${\rm CV}(g)=(ng)^{-1}R(L)+\{n(n-1)\}^{-1}\sum_{i\neq j}\{(1-n^{-1})(L*L)_g-2L_g\}(X_i-X_j),$$
	is the cross-validation criterion function associated to the kernel density estimator
	\begin{equation} \label{kde}
		\widehat f(x;g)=n^{-1}\sum_{j=1}^n L_g(x-X_j),
	\end{equation}
	where we denote $R(\alpha)=\int_{\mathbb R} \alpha(x)^2 dx$ for an arbitrary square integrable real function $\alpha$,
	$L$ is a kernel, that is, a bounded, symmetric and integrable function with unit integral, $*$ denotes the convolution product,
	and $L_g(x)=L(x/g)/g$ is the scaled kernel associated to $L$.
	
	The rest of this appendix is organized as follows.
	In Section \ref{s.existence} we provide mild conditions on the kernel and the density that ensure the existence of a bandwidth $g_{\MSE}$, called exact optimal bandwidth, that minimizes the mean squared error ${\rm MSE}(g)$ of $\widecheck\psi(g)$.
	In Section \ref{s.limit} we study the asymptotic properties of this bandwidth.
	In Section \ref{s.order} we prove that $g_{\MSE}$ is asymptotically equivalent to the bandwidth $g_{\MISE}$
	that minimizes the mean integrated squared error of the kernel density estimator (\ref{kde}), and we
	obtain the relative rates of convergence of $g_{\MSE}$ to $g_{\MISE}$.
	These results suggest that using a kernel $L$ of order higher than 2 may not be convenient as the quality of the approximation between
	$g_{\MSE}$ and $g_{\MISE}$ decreases when the kernel order increases.
	All the proofs are deferred to Section \ref{s.proofs}.

	\subsection{Existence of an exact optimal bandwidth} \label{s.existence}
	
	For $g>0$, we have
	$$\widecheck\psi(g) = -(ng)^{-1}R(L) + U_{M_g} + n^{-1} U_{N_g},$$
	where $M_g$ and $N_g$ are the scaled kernels associated to
	$$M=2L-L*L \quad \mbox{and} \quad N=L*L,$$
	respectively, and for a given bounded, symmetric and integrable function $\varphi$, $U_{\varphi_g}$ is the $U$-statistic
	$$U_{\varphi_g} = \frac{2}{n(n-1)} \sum_{1 \leq i < j \leq n} \varphi_g(X_i - X_j).$$
	
	Taking into account that
	$\E U_{\varphi_g} = R_{\varphi,g}(f),$
	with
	$$R_{\varphi,g}(f) = \int_{\mathbb R} \varphi_g*f(x) f(x) dx,$$
	we deduce that the bias ${\rm B}(g)$ of $\widecheck\psi(g)$ can be written as
	\begin{equation}  \label{Bg}
		{\rm B}(g)  = \E \widecheck\psi(g) - \psi = - (ng)^{-1}R(L) + R_{M,g}(f) + n^{-1} R_{N,g}(f) - \psi. 
	\end{equation}
	From this equality and Equation (10) in \citet[][p.~296]{ChaMNP:07} we see that
	${\rm B}(g) = - \MISE(g)$, where $\MISE(g) = \E \int_{\mathbb R} \{ \widehat f(x;g) - f(x) \}^2 dx$ is the mean integrated squared error of the kernel density estimator (\ref{kde}).
	
	On the other hand, from the covariance formula between two $U$-statistics \citep[see][Theorem 2, p.~17]{Lee:90}, we deduce that the variance ${\rm V}(g)$ of $\widecheck\psi(g)$ is given by
	\begin{align}
		{\rm V}(g) & = \Var(U_{M_g}) + 2 n^{-1} \Cov(U_{M_g}, U_{N_g} ) + n^{-2} \Var(U_{N_g})  \nonumber \\
		& =  4(n-2)\{n(n-1)\}^{-1} \big\{ T_{M,M,g}(f) + 2 n^{-1} T_{M,N,g}(f) + n^{-2} T_{N,N,g}(f) \big\} \nonumber \\
		& \quad - (4n-6)\{n(n-1)\}^{-1} \big\{ R_{M,g}(f)^2 + 2 n^{-1} R_{M,g}(f)R_{N,g}(f) + n^{-2} R_{N,g}(f)^2 \big\} \nonumber \\
		& \quad + 2g^{-1}\{n(n-1)\}^{-1} \big\{ R_{M^2\!,g}(f) + 2 n^{-1} R_{M\cdot N,g}(f) + n^{-2} R_{N^2\!,g}(f) \big\}, \label{Vg}
	\end{align}
	with
	$$T_{\varphi,\phi,g}(f) = \int_{\mathbb R} \varphi_g*f(x)\phi_g*f(x) f(x) dx,$$
	where $\varphi$ and $\psi$ are bounded, symmetric and integrable functions.
	
	Combining equations (\ref{Bg}) and (\ref{Vg}) we obtain the following exact formula for the mean squared error $\MSE(g)$ of the estimator
	$\widecheck\psi(g)$:
	\begin{align}
		\MSE(g) & = \big\{ (ng)^{-1}R(L) - R_{M,g}(f) - n^{-1} R_{N,g}(f) + \psi \big\}^2 \nonumber \\
		& \quad + 4(n-2)\{n(n-1)\}^{-1} \big\{ T_{M,M,g}(f) + 2 n^{-1} T_{M,N,g}(f) + n^{-2} T_{N,N,g}(f) \big\} \nonumber \\
		& \quad - (4n-6)\{n(n-1)\}^{-1} \big\{ R_{M,g}(f)^2 + 2 n^{-1} R_{M,g}(f)R_{N,g}(f) + n^{-2} R_{N,g}(f)^2 \big\} \nonumber \\
		& \quad + 2g^{-1}\{n(n-1)\}^{-1} \big\{ R_{M^2\!,g}(f) + 2 n^{-1} R_{M\cdot N,g}(f) + n^{-2} R_{N^2\!,g}(f) \big\}. \label{MSEg}
	\end{align}
	This exact error formula is the analogue of formula (5) in \citet[][p.~526]{ChaT:12} and will be useful to explore the existence and limit behavior of the optimal bandwidth.
	
	In the following results we will make the next assumptions on the kernel $L$ and the density $f$:
	
	\medskip
	
	\noindent
	(L1)
	$L$ is a bounded, symmetric and integrable function with unit integral, which is continuous at zero, with $R(L)<2L(0)$.
	
	\medskip
	
	\noindent
	(D1)
	$f$ is bounded.
	
	\medskip
	
	The next result shows that under mild conditions there is always an exact
	optimal bandwidth, that is, a bandwidth which minimizes the exact MSE of estimator $\widecheck\psi(g)$.
	It is the analogue of Theorem 1 in \citet[][p.~526]{ChaT:12} for the `diagonals-in' kernel estimator
	$\widetilde \psi_{\rm D}(g)=n^{-2}\sum_{i,j=1}^nL_g(X_i-X_j)$.
	
	\begin{theorem} \label{existsexact}
		Under assumptions {\rm (L1)} and {\rm (D1)},
		there exists $g_{\MSE} = g_{\MSE,n}(f)$ such that $\MSE(g_{\MSE}) \leq \MSE(g)$, for all $g>0$.
	\end{theorem}

	\subsection{Limit behavior of $g_{\MSE}$} \label{s.limit}
	
	From formula (\ref{MSEg}) and Lemma \ref{lemmaRT} in Section \ref{s.proofs} below it follows
	that ${\rm MSE}(g)\to 0$ for any bandwidth sequence $g = g_n$ such that $g\to 0$ and
	$ng \to \infty$, as $n\to\infty$. Therefore, conditions $g \to 0$ and $ng \to\infty$ are
	sufficient for $\widecheck\psi(g)$ to be consistent. It is natural, then, to wonder if the bandwidth
	$g_\mathrm{MSE}$ also fulfill the previous consistency conditions. We will see
	that the second condition holds quite generally but the same is not necessarily true for the
	first one. This is similar to the situation with the exact optimal bandwidth for the `diagonals-in' kernel estimator
	$\widetilde \psi_{\rm D}(g)$, as shown in \citet{ChaT:12}.
	
	\begin{theorem} \label{limitngMSE}
		Under assumptions {\rm (L1)} and {\rm (D1)}, we have
		$ng_{\rm MSE} \to \infty$, as $n\to \infty$.
	\end{theorem}
	
	For the analysis of the limit behavior of the sequence $g_{\MSE}$ we use the notation $\varphi_K(t)=\int_{\mathbb R} e^{itx}K(x)
	dx,\, t\in \mathbb{R}$, for the characteristic function of an integrable
	function $K$, and for every density $f$ and every kernel $L$, we denote
	\begin{eqnarray*}
		C_f & = & \sup \{ r \geq 0 : \varphi_f(t) \neq 0 \; \mbox{ a.e. for } t\in [0,r] \}, \\
		D_f & = & \sup \{ t \geq 0 : \varphi_f(t) \neq 0 \}, \\
		S_L & = & \inf \{ t \geq 0 : \varphi_L(t) \neq 1 \} \; \mbox{and} \\
		T_L & = & \inf \{ r \geq 0 : \varphi_L(t) \neq 1 \; \mbox{ a.e. for
		} t\geq r \}.
	\end{eqnarray*}
	A detailed discussion about these quantities is presented in \citet{ChaMNP:07}. In
	particular, we note that all these exist, with $C_f,D_f$ possibly being infinite, $S_L,T_L \in
	[0,\infty)$, $C_f \leq D_f$ and $S_L\leq T_L$. By definition, $S_L>0$ for superkernels
	and $S_L=0$ if $L$ is a kernel of finite order $\nu$ (even), that is, if $m_j(L)=0$ for $j=1,2,\dots,\nu-1$ and $m_\nu(L)\neq0$ with $|m_\nu|(L)<\infty$, where $m_j(L)=\int_\mathbb{R} u^j L(u) du$  and $|m_j|(L)=\int_\mathbb{R} |u^j L(u)| du$ \citep[see][]{ChaMN:07}.
	
	An additional assumption on the kernel is needed to show that $g_\mathrm{MSE}$ converges to zero:
	
	\medskip
	
	\noindent
	(L2)
	$L$ is such that $\varphi_L(t) \leq 1$, for all $t\in \mathbb R$.
	
	\medskip
	
	In the following result we show that $g_\mathrm{MSE}$ converges to zero under very general conditions. In
	particular, if $L$ is a kernel of finite order, the convergence to zero takes
	place with no additional conditions on $f$ other than being bounded. The same property occurs in the
	superkernel case whenever the characteristic function of $f$ has unbounded support.
	
	\begin{theorem} \label{limitgMSE}
		Under assumptions {\rm (L1)}, {\rm (L2)} and {\rm (D1)},
		if $S_L=0$ or $D_f=\infty$ then $g_{\MSE} \rightarrow 0$, as $n\to\infty$.
	\end{theorem}
	
	\subsection{The bandwidths $g_{\MSE}$ and $g_{\MISE}$} \label{s.order}
	
	In order to study the order of convergence to zero of the exact optimal bandwidth $g_{\MSE}$,
	we need some additional assumptions on the kernel $L$ and the density $f$:
	
	\medskip
	
	\noindent
	(L3)
	$L$ is a kernel of finite order $\nu$ (even) such that $(-1)^{\nu/2}m_\nu(L) < 0$ and $|m|_{2\nu}(L)<\infty$.
	
	\smallskip
	
	\noindent
	(D2)
	$f$ has bounded and integrable derivatives up to order $\nu$.
	
	\medskip
	
	Under conditions (L1)--(L3), (D1) and (D2), from Lemma \ref{asymRT} and the fact that $M=2L-L*L$ and $N=L*L$ are kernels of orders $2\nu$ and $\nu$, respectively, with $m_{2\nu}(M)=-(2\nu)!m_\nu(L)^2/(\nu!)^2$ and $m_{\nu}(N)=2m_\nu(L)$, we conclude that
	the bias and variance of $\widecheck\psi(g)$ given by equations (\ref{Bg}) and (\ref{Vg}), respectively, admit the asymptotic expansions, as $g \to 0$,
	\begin{equation} \label{Bgexpansion}
		{\rm B}(g) = - (ng)^{-1} R(L) - g^{2\nu} R(f^{(\nu)}) m_\nu(L)^2/(\nu!)^2 + O(n^{-1}) + o(g^{2\nu})
	\end{equation}
	and
	$${\rm V}(g) = 4n^{-1} \Var f(X_1) + O\big( n^{-1}g^{2\nu} + n^{-2} g^{-1} \big),$$
	from which we get the following asymptotic expansion for the mean square error of $\widecheck\psi(g)$:
	\begin{align}
		\MSE(g) & = 4n^{-1} \Var f(X_1) + \big\{ (ng)^{-1} R(L) + g^{2\nu} R(f^{(\nu)}) m_\nu(L)^2/(\nu!)^2 \big\}^2 \nonumber \\
		& \quad + O(n^{-2}g^{-1}) + o(n^{-1} g^{2\nu-1} + g^{2\nu} ). \label{expMSE}
	\end{align}
	
	Based on these asymptotic expansions, in the following result we start by establishing that the exact optimal bandwidth $g_{\MSE}$ is of order $n^{-1/(2\nu+1)}$. Therefore, for $g$ of order $n^{-1/(2\nu+1)}$ and the fact that ${\rm B}(g) = - \MISE(g)$, we get the equality
	$\MSE(g) = 4n^{-1} \Var f(X_1) + \MISE(g)^2 (1+o(1))$,
	that suggests that the optimal bandwidth $g_{\MSE}$ may be asymptotically equivalent to the bandwidth $g_{\MISE}$
	that minimizes the mean integrated squared error of the kernel density estimator $\widehat f(\cdot;g)$ given by (\ref{kde})
	(regarding the existence and asymptotic behaviour of $g_{\MISE}$, see \citealp{ChaMNP:07}).
	This fact, together with the order of convergence of the relative error $g_{\MSE}/g_{\MISE}-1$, is established in the following result.

	\begin{theorem} \label{ordergMSE}
		Under assumptions {\rm (L1)}--{\rm (L3)}, {\rm (D1)} and {\rm (D2)}, we have:
		
		\smallskip
		
		\noindent
		(a) The bandwidths $g_{\MSE}$ and $g_{\MISE}$ are of the same order, that is,
		$$0 < \liminf n^{1/(2\nu+1)} g_{\MSE} \leq \limsup n^{1/(2\nu+1)} g_{\MSE} < \infty.$$
		
		\noindent
		(b) The bandwidths $g_{\MSE}$ and $g_{\MISE}$ are asymptotically equivalent, that is,
		$$g_{\MSE}/g_{\MISE} \to 1.$$
		
		\noindent
		(c) There exists a constant $C$, depending on $L$ and $f$, such that
		$$n^{1/(2\nu+1)} \big( g_{\MSE}/g_{\MISE} -1 \big) = C (1+o(1)).$$
		
	\end{theorem}

	This result supports the idea of taking $g_{\rm MISE}$ as the target bandwidth for $\widecheck\psi(g)$,
	giving us an additional motivation for estimating $\psi$
	by $\widehat\psi = \widecheck\psi(\widehat{g}_{\rm CV})$.
	Moreover, it also enables us to recommend the use of a kernel $L$ of second order ($\nu=2$) because the order of convergence to zero of the relative error $g_{\MSE}/g_{\MISE} -1$ is a decreasing function of the kernel order.
	
	\subsection{Proofs} \label{s.proofs}
	
	We start by establishing the continuity and limit behaviour of the functions $g \mapsto R_{\varphi,g}(f)$ and $g \mapsto T_{\varphi,\phi,g}(f)$ defined in Section \ref{s.existence}.
	
	\begin{lemma} \label{lemmaRT}
		Under assumption {\rm (D1)}, assume that $\varphi, \phi$ are bounded and integrable functions.
		
		\medskip
		
		\noindent
		(a) The function $g \mapsto R_{\varphi,g}(f)$ is continuous with
		$$\lim_{g \to 0} R_{\varphi,g}(f) = R(f) \int_{\mathbb R} \varphi(u) du.$$
		In addition, if
		$\varphi$ is continuous at zero, then
		$\lim_{g \to \infty} g\,R_{\varphi,g}(f) = \varphi(0)$.
		
		\noindent
		(b) The function $g \mapsto T_{\varphi,\phi,g}(f)$ is continuous with
		$$\lim_{g \to 0} T_{\varphi,\phi,g}(f) = R(f^{3/2}) \int_{\mathbb R} \varphi(u) du \int_{\mathbb R} \phi(u) du.$$
		In addition, if $\varphi$ and $\phi$ are continuous at zero, then
		$\lim_{g \to \infty} g^2 \, T_{\varphi,\phi,g}(f) = \varphi(0)\phi(0)$.
		
	\end{lemma}
	
	\noindent
	{\it Proof:}
	Taking into account that
	\begin{equation} \label{newexpR}
		R_{\varphi,g}(f) = \int_{\mathbb R} \varphi(u) (\bar{f}*f)(ug) du,
	\end{equation}
	where $\bar{f}(u)=f(-u)$, part (a) follows from the dominated convergence theorem and
	the boundedness and the continuity of the convolution product of square integrable functions
	\citep[see also][Lemma, p.~296]{ChaMNP:07}.
	Moreover, we have
	\begin{equation} \label{newexpT}
		T_{\varphi,\phi,g}(f) = \int_{\mathbb R}\!\int_{\mathbb R} \varphi(u)\phi(v)
		\big(\bar{f} \odot \bar{f} \odot f \big)(ug,vg)dudv,
	\end{equation}
	where we are denoting
	$$(\alpha \odot\beta \odot\gamma)(x,y)=\int_{\mathbb R} \alpha(x-z)\beta(y-z)\gamma(z)dz.$$
	Therefore, part (b) follows from the dominated convergence theorem and
	the boundedness and the continuity of $\alpha\odot\beta\odot\gamma$ when $\alpha,\beta$ and $\gamma$ are bounded and integrable functions \citep[see][proof of Lemma 1, p.~539]{ChaT:12}. \hfill$\blacksquare$

	\bigskip
	
	\noindent
	{\it Proof of Theorem \ref{existsexact}:}
	From the expression (\ref{MSEg}) for the MSE function, and the properties of the functions $g \mapsto R_{\varphi,g}(f)$ and
	$g \mapsto T_{\varphi,\phi,g}(f)$ shown in Lemma \ref{lemmaRT}, we conclude that $g \mapsto \MSE(g)$ is a continuous function such that
	$\lim_{g \to 0} \MSE(g) = \infty$ and $\lim_{g \to \infty} \MSE(g) = \psi^2$. Moreover, from the hypotheses on $L$ we have
	$$\lim_{g \to \infty} g^2 ( \MSE(g) - \psi^2) = - \infty,$$
	which enables us to conclude that we can choose $g>0$ big enough so that $\MSE(g) < \psi^2$. This concludes the proof. \hfill$\blacksquare$
	
	\bigskip
	
	\noindent
	{\it Proof of Theorem \ref{limitngMSE}:}
	Suppose that $ng_{\MSE}$ does not converge to infinity. Then $ng_{\MSE}$ has a subsequence which is upper
	bounded by some positive constant $C$. Therefore, along that subsequence we have $g_{\MSE} \to 0$.
	From (\ref{Bg}) this implies that
	$$\limsup_{n \to \infty} \MSE(g_{\MSE}) \geq \limsup_{n \rightarrow \infty} {\rm B}(g_{\MSE})^2
	= \limsup_{n \to \infty} \{n^{-1}g_{\MSE}^{-1}R(L)\}^2 \geq \{R(L)/C\}^2 >0,$$
	which contradicts the fact that $\lim_{n \to \infty} \MSE(g_{\MSE})=0$, as we can deduce from
	$0 \leq \MSE(g_{\MSE}) \leq \MSE(g) \rightarrow 0$, for any bandwidth sequence $g = g_n$ such that $g\to 0$ and
	$ng \to \infty$, as $n\to\infty$. \hfill$\blacksquare$

	\bigskip
	
	\noindent
	{\it Proof of Theorem \ref{limitgMSE}:}
	Denote by $\Lambda_{f,L}$ the set of accumulation points of the sequence $(g_{\MSE})$. Take
	$0<\lambda\in \Lambda_{f,L}$ and $(g_{n_k})$ a subsequence of
	$(g_{\MSE})$ such that $\lambda=\lim_{k\to\infty} g_{n_k}$.
	Writing ${\rm B}(g;n)$ and $\MSE(g;n)$ for ${\rm B}(g)$ and
	$\MSE(g)$, respectively, from equalities (\ref{Bg}) and (\ref{Vg}) we get that, for fixed $g>0$,
	$$\lim_{n\to\infty} \MSE(g;n) = \lim_{n\to\infty} {\rm B}(g;n)^2 = \{R_{M,g}(f) - \psi \}^2,$$ so
	that using Lemma \ref{lemmaRT} and Theorem \ref{limitngMSE}, we
	obtain
	\begin{align*}
		0 & = \lim_{g\to 0} \{ R_{M,g}(f)-\psi \}^2
		=  \lim_{g\to 0} \lim_{k\to\infty} \MSE(g;n_k) \\
		& \geq \lim_{k\to\infty} \MSE(g_{n_k};n_k) \geq \lim_{k\to\infty} {\rm B}(g_{n_k};n_k)^2
		=  \{ R_{M,\lambda}(f) - \psi \}^2.
	\end{align*}
	Therefore
	$$\Lambda_{f,L} \subset \{ \lambda \geq 0 : R_{M,\lambda}(f) = \psi \}$$
	Taking into account that $\varphi_M(t) = (2-\varphi_L(t))\varphi_L(t) \leq 1$, for all $t \in \mathbb{R}$,
	then reasoning as in the proof of Theorem 4 in \citet{ChaT:12} we conclude that
	$$\Lambda_{f,L} \subset \left[0, \min\left(\frac{S_M}{C_f},\frac{T_M}{D_f}\right) \right].$$
	As $S_M=S_L$ and $T_M=T_L$, we
	finally get
	$$0 \leq \limsup_{n\to\infty} g_{\MSE} \leq
	\min\left(\frac{S_L}{C_f},\frac{T_L}{D_f}\right),$$
	which concludes the proof. \hfill$\blacksquare$
	
	\medskip
	
	In the following result we study the differentiability of the functions $g \mapsto R_{\varphi,g}(f)$
	and $g \mapsto T_{\varphi,\phi,g}(f)$ defined in Section \ref{s.existence},
	and we provide asymptotic expansions for them and the corresponding derivatives when $g \to 0$.
	
	\begin{lemma} \label{asymRT}
		Under assumptions {\rm (D1)} and {\rm (D2)}, let $\varphi$ and $\phi$ be bounded and integrable functions
		such that $|m|_{\ell}(\varphi), |m|_{\ell}(\phi)<\infty$ and
		$m_k(\varphi)=m_k(\phi)=0$, for all $k=1,\dots,\ell-1$, for some even integer $\ell$ such that $2\leq \ell \leq 2\nu$.
		
		\medskip
		
		\noindent
		(a) The function $g \mapsto R_{\varphi,g}(f)$ is twice differentiable with
		$$R_{\varphi,g}(f) = R(f) \, m_{0}(\varphi)
		+ (-1)^{\ell/2} g^{\ell} R(f^{(\ell/2)}) \, m_{\ell}(\varphi) / \ell! + o(g^{\ell}),$$
		$$dR_{\varphi,g}(f)/dg = (-1)^{\ell/2} g^{\ell-1} R(f^{(\ell/2)}) \, m_{\ell}(\varphi) / (\ell-1)! + o(g^{\ell-1})$$
		and
		$$d^2R_{\varphi,g}(f)/dg^2 = (-1)^{\ell/2} g^{\ell-2} R(f^{(\ell/2)}) \, m_{\ell}(\varphi) / (\ell-2)! + o(g^{\ell-2}).$$
		
		\noindent
		(b) The function $g \mapsto T_{\varphi,\phi,g}(f)$ is differentiable with
		$$T_{\varphi,\phi,g}(f) = \eta_{0} m_{0}(\varphi) m_{0}(\phi) +
		(-1)^{\ell/2} g^{\ell} \eta_{\ell} \, \big( m_\ell(\varphi) m_0(\varphi) + m_0(\varphi) m_\ell(\phi) \big) / \ell! + o(g^{\ell})$$
		and
		$$dT_{\varphi,\phi,g}(f)/dg =
		(-1)^{\ell/2} g^{\ell-1} \eta_{\ell} \, \big( m_\ell(\varphi) m_0(\varphi) + m_0(\varphi) m_\ell(\phi) \big) / \ell! + o(g^{\ell-1}),$$
		where $\eta_{\ell}= \int_\mathbb{R} f^{(\ell/2)}(z) (f^2)^{(\ell/2)}(z)\, dz$.
	\end{lemma}
	
	\noindent
	\textit{Proof:}
	Taking into account assumption (D2), the functions $\bar{f}*f$ and $h=\bar{f} \odot \bar{f}\odot f$ have
	continuous derivatives up to order $2\nu$.
	From the expressions (\ref{newexpR}) and (\ref{newexpT}) for $R_{\varphi,g}(f)$ and $T_{\varphi,\phi,g}(f)$, respectively,
	standard Taylor's expansions and the dominated convergence theorem lead to
	\begin{align*}
		R_{\varphi,g}(f)
		& = \sum_{i=0}^{\ell-1} \frac{g^i}{i!} \, (\bar{f}*f)^{(i)}(0) \int_\mathbb{R} u^i \varphi(u) du \\
		& \quad + \frac{g^{\ell}}{(\ell-1)!} \int_\mathbb{R} \! \int_0^1 u^\ell \varphi(u) (1-t)^{\ell-1} (\bar{f}*f)^{(\ell)} (ugt) dt du \\
	& = R(f) \, m_{0}(\varphi) + g^{\ell} (\bar{f}*f)^{(\ell)}(0) \, m_{\ell}(\varphi) / \ell! + o(g^{\ell})
\end{align*}
and
\begin{align*}
	\lefteqn{T_{\varphi,\phi,g}(f) } \\
	& = \sum_{0 \leq i+j \leq \ell-1} \frac{g^{i+j}}{i!j!} \, \frac{\partial^{i+j} h}{\partial x^i \partial y^{j}}(0,0)
	\int_\mathbb{R} u^i \varphi(u) du \int_\mathbb{R} v^j \phi(v) dv \\
	& \quad + \sum_{i+j=\ell} \frac{\ell g^\ell}{i!j!} \int_\mathbb{R}\!\int_\mathbb{R}\!\int_0^1 u^i \varphi(u)v^j\phi(v) (1-t)^{\ell-1}
	\frac{\partial^{\ell} h}{\partial x^i \partial y^{j}} (tug,tvg) dtdudv \\
	& = \eta_0 \, m_{0}(\varphi) m_{0}(\phi) + \frac{g^{\ell}}{\ell!} \bigg(
	m_\ell(\varphi) m_0(\varphi) \frac{\partial^{\ell} h}{\partial x^{\ell}} (0,0)
	+ m_0(\varphi) m_\ell(\phi) \frac{\partial^{\ell} h}{\partial y^{\ell}} (0,0) \bigg) + o(g^{\ell}).
\end{align*}

Moreover, from the differentiation theorem under the integral sign we have
$$dR_{\varphi,g}(f)/dg = \int_\mathbb{R} u\varphi(u) (\bar{f}*f)^\prime(ug) du,$$
$$d^2R_{\varphi,g}(f)/dg^2 = \int_\mathbb{R} u^2\varphi(u) (\bar{f}*f)^{\prime\prime}(ug) du$$
and
$$dT_{\varphi,\psi,g}(f)/dg = \int_{\mathbb R}\!\int_{\mathbb R} \varphi(u)\phi(v)
\bigg( u \frac{\partial h}{\partial x} (ug,vg)
+ v \frac{\partial h}{\partial y} (ug,vg) \bigg) dudv.$$
By using again Taylor's expansions and the dominated convergence theorem we get
\begin{align*}
	dR_{\varphi,g}(f)/dg
	& = \frac{g^{\ell-1}}{(\ell-1)!} (\bar{f}*f)^{(\ell)}(0) \, m_\ell(\varphi)  + o( g^{\ell-1} ),
\end{align*}
\begin{align*}
	d^2R_{\varphi,g}(f)/dg^2
	& = \frac{g^{\ell-2}}{(\ell-2)!} (\bar{f}*f)^{(\ell)}(0) \, m_\ell(\varphi)  + o( g^{\ell-2} ),
\end{align*}
and
\begin{align*}
	dT_{\varphi,\phi,g}(f)/dg
& =   \frac{g^{\ell-1}}{(\ell-1)!} \bigg(
m_\ell(\varphi) m_0(\varphi) \frac{\partial^{\ell} h}{\partial x^{\ell}} (0,0)
+ m_0(\varphi) m_\ell(\phi) \frac{\partial^{\ell} h}{\partial y^{\ell}} (0,0) \bigg) + o(g^{\ell-1}).
\end{align*}

Parts (a) and (b) follow now from the fact that
$(\bar{f}*f)^{(\ell)} (0) = (-1)^{\ell/2} R(f^{(\ell/2)})$
and $\frac{\partial^{\ell} h}{\partial x^{\ell}} (0,0)=\frac{\partial^{\ell} h}{\partial y^{\ell}} (0,0)
= (-1)^{\ell/2} \int_\mathbb{R} f^{(\ell/2)}(z) (f^2)^{(\ell/2)}(z) dz$. \hfill$\blacksquare$

\bigskip

\noindent
{\it Proof of Theorem \ref{ordergMSE}:}
(a) From expansion (\ref{expMSE}) and taking for $g$ the bandwidth $g=c\, n^{1/(2\nu+1)}$, with $c>0$, we get
$$\lim n^{4\nu/(2\nu+1)} \left( \MSE(g_0)-4n^{-1} \Var f(X_1) \right) = U(c),$$
with
$$U(c) = \{ c^{-1} R(L) + c^{2\nu} R(f^{\nu}) m_\nu(L)^2 / (\nu!)^2 \}^2.$$
Therefore, as ${\rm MSE}(g_{\MSE})\leq \MSE(g_0)$, we have
\begin{equation} \label{limsupfinite}
\limsup n^{4\nu/(2\nu+1)} \left( \mathrm{MSE}(g_{\MSE})-4n^{-1} \Var f(X_1) \right) \leq U(c) < \infty.
\end{equation}
Moreover, using the fact that $g_{\MSE} \to 0$, from expansion (\ref{expMSE})
we also get
\begin{eqnarray*}
\lefteqn{n^{4\nu/(2\nu+1)} \left( \mathrm{MSE}(g_{\MSE})-4n^{-1} \Var f(X_1) \right) } \\
& = & \left\{ (n^{1/(2\nu+1)} g_{\MSE})^{-1} R(L)
+  (n^{1/(2\nu+1)} g_{\MSE})^{2\nu} R(f^{(\nu)}) m_\nu(L)^2/(\nu!)^2 \right\}^2 \\
& & + o\left( (n^{1/(2\nu+1)} g_{\MSE})^{-1}+(n^{1/(2\nu+1)} g_{\MSE})^{2\nu-1}  + (n^{1/(2\nu+1)} g_{\MSE})^{4\nu} \right),
\end{eqnarray*}
which contradicts (\ref{limsupfinite}) if $\liminf n^{1/(2\nu+1)} g_{\MSE}=0$
or $\limsup n^{1/(2\nu+1)} g_{\MSE} = \infty$. This completes the proof of part (a). 

\smallskip

\noindent
(b) Taking into account that inequality (\ref{limsupfinite}) is true for all $c>0$, we also have
\begin{equation} \label{limsupfinite1}
\limsup n^{4\nu/(2\nu+1)} \left( \mathrm{MSE}(g_{\MSE})-4n^{-1} \Var f(X_1) \right) \leq U(c_0(f,L)),
\end{equation}
where
\begin{equation} \label{c0fL}
c_0(f,L) = \bigg( \frac{R(L) (\nu!)^2}{2\nu R(f^{(\nu)}) m_\nu(L)^2} \bigg)^{1/(2\nu+1)}
\end{equation}
is the minimizer of the function $c \mapsto U(c)$. By using standard arguments, inequality (\ref{limsupfinite1})
enables us to conclude that $c_0(f,L)$ is the only accumulation point of the bounded sequence $(n^{1/(2\nu+1)} g_{\MSE})$, that is,
$g_{\MSE}$ is asymptotically equivalent to $g_0 = c_0(f,L)\,n^{-1/(2\nu+1)}$ \citep[see][pp.~44--46]{Cha:04}. This concludes the proof of part (b) as it is well-known that the sequence $g_0$ is asymptotically equivalent to $g_{\MISE}$ \citep[see][pp.~293--294]{ChaMNP:07}.

\smallskip

\noindent
(c) From Lemma \ref{asymRT} and equality (\ref{Bg}) the function ${\rm B}(g)$ is twice differentiable with
\begin{align}
{\rm B}^{\prime\prime}(g)
& = - 2 n^{-1} g^{-3} R(L) - 2\nu (2\nu-1) g^{2\nu-2} R(f^{(\nu)})  m_\nu(L)^2/(\nu!)^2 \nonumber \\
& \quad + O(n^{-1} g^{\nu-2}) + o(g^{2\nu-2}). \label{d2Bexpansion}
\end{align}
Therefore, a Taylor's expansion leads to
$$0=\MISE^\prime(g_{\MISE}) = - {\rm B}^\prime(g_{\MISE}) = - {\rm B}^\prime(g_{\MSE})
- (g_{\MISE}-g_{\MSE}) {\rm B}^{\prime\prime}(\widetilde{g}),$$
from which we deduce that
\begin{equation} \label{exp1}
n^{1/(2\nu+1)} \big( g_{\MSE}/g_{\MISE} -1 \big) = c_0(f,L) \, n g_{\MSE} {\rm B}^\prime(g_{\MSE})
\big( n \widetilde{g}^3 {\rm B}^{\prime\prime}(\widetilde{g}) \big)^{-1} (1+o(1)),
\end{equation}
for some $\widetilde{g}$ between $g_{\MISE}$ and $g_{\MSE}$, and $c_0(f,L)$ given by (\ref{c0fL}).
Taking into account that $\widetilde{g}$ is asymptotically equivalent to $g_{\MSE}$,
from equation (\ref{d2Bexpansion}) we obtain
\begin{equation} \label{exp2}
n \widetilde{g}^3 {\rm B}^{\prime\prime}(\widetilde{g}) = -(2\nu + 1) R(L) (1 + o(1)).
\end{equation}
On the other hand, from Lemma \ref{asymRT} and equality (\ref{Vg}) we know that the function ${\rm V}(g)$ is differentiable with
\begin{align*}
{\rm V}^\prime(g)
& = - 2 n^{-2} g^{-2} R(f) R(M) - 8 n^{-1} g^{2\nu-1} \{ \eta_{2\nu} - 2\nu R(f^{(\nu)}) R(f) \} m_\nu(L)^2/(\nu!)^2  \\
& \quad + o( n^{-2} g^{-2} + n^{-1} g^{2\nu-1})
\end{align*}
and
$$n^2g_{\MSE}^2 {\rm V}^\prime(g_{\MSE}) = D(f,L) (1+o(1)),$$
with $D(f,L)=-2R(f)(R(M)-4R(L))-4\nu^{-1}\eta_{2\nu} R(f^{(\nu)})^{-1} R(L)$.
Therefore, by using equation $\MSE^\prime(g_{\MSE}) = 2{\rm B}(g_{\MSE}){\rm B}^\prime(g_{\MSE}) + {\rm V}^\prime(g_{\MSE}) = 0$,
and the fact that $ng_{\MSE} {\rm B}(g_{\MSE}) = -(2\nu +1)(2\nu)^{-1}R(L)(1+o(1)),$
which can be deduce from equation (\ref{Bgexpansion}), we get
\begin{equation} \label{exp3}
ng_{\MSE} {\rm B}^\prime(g_{\MSE}) = \nu (2\nu+1)^{-1} D(f,L) R(L)^{-1} (1+o(1)).
\end{equation}
The stated result with $C=-\nu(2\nu+1)^{-2} c_0(f,L)D(f,L) R(L)^{-2}$ follows now from (\ref{exp1}), (\ref{exp2}) and (\ref{exp3}). \hfill$\blacksquare$

\section{Additional information on the simulation study}\label{app:b}

Here, a more detailed account of the simulation results summarized in Section 3 of the main text is given. Let us recall that  the test densities included in the study are the 15 normal mixture densities introduced in \cite{MW92}, plus the 10-modal normal mixture described in \cite{L99}, which will be referred to as Density \#16. 

These test densities are all not equally hard to estimate. To measure how easy is a given density $f$ to estimate we can resort to the functional
$$Q(f)=\inf_{u>0}u^{-1}\int_{\mathbb R}f(x)^{1/2}\rho\{u^5f''(x)f(x)^{-1/2}\}dx$$
introduced in \cite{WD93}, with $\rho(t)=\mathbb E|Z-t|$, where $Z$ is a random variable with a standard normal distribution. These authors also showed that this functional is approximately minimised for the Beta$(5.3,5.3)$ distribution, yielding a value of about 1.92. The value of $Q(f)$ for each of the test densities in the simulation study is given in Table \ref{tab:Qf}.

\begin{table}[h]
\centering
\begin{tabular}{|l||c|c|c|c|c|c|c|c|}\hline
Density \#&1&2&3&4&5&6&7&8\\\hline
$Q(f)$& 1.99& 2.16& 4.36& 4.2& 3.26& 2.29& 2.59& 2.55\\\hline\hline
Density \#&9&10&11&12&13&14&15&16\\\hline
$Q(f)$& 2.62& 4.18& 7.08& 5.11& 3.82& 6.2& 5.32& 4.99\\\hline
\end{tabular}
\caption{Density-estimation-difficulty functional $Q(f)$.}
\label{tab:Qf}
\end{table}

From Table \ref{tab:Qf}, the test densities can be categorised into three groups: the first one comprises densities 1, 2, 6, 7, 8 and 9, which can be considered as easy-to-estimate densities, with $Q(f)<2.7$; densities 3, 4, 5, 10 and 13 could be seen as medium-estimation-difficulty densities, since they have $3.2<Q(f)<4.4$; and densities 11, 12, 14, 15 and 16 may be cast as difficult-to-estimate densities, with $Q(f)>4.9$.

\begin{figure}[!pt]
\centering
\begin{tabular}{cccc}
\quad Density \#1&\quad Density \#2 &\quad Density \#3&\quad  Density \#4\\
\includegraphics[width=0.23\textwidth]{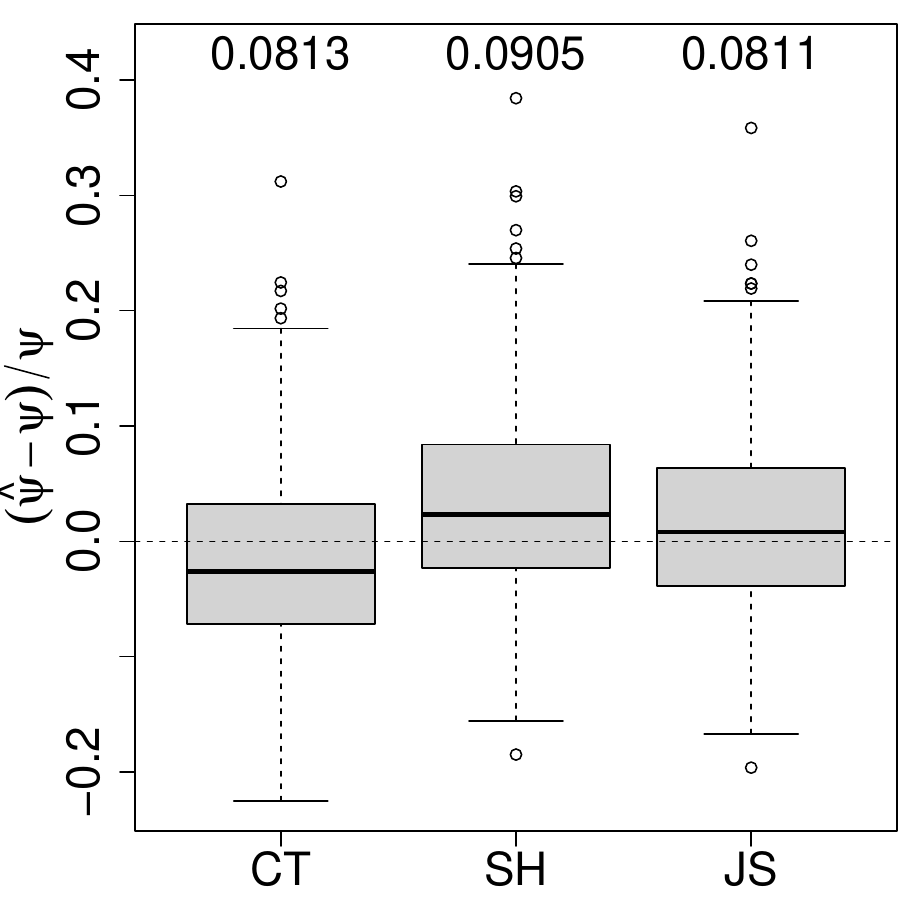} &
\includegraphics[width=0.23\textwidth]{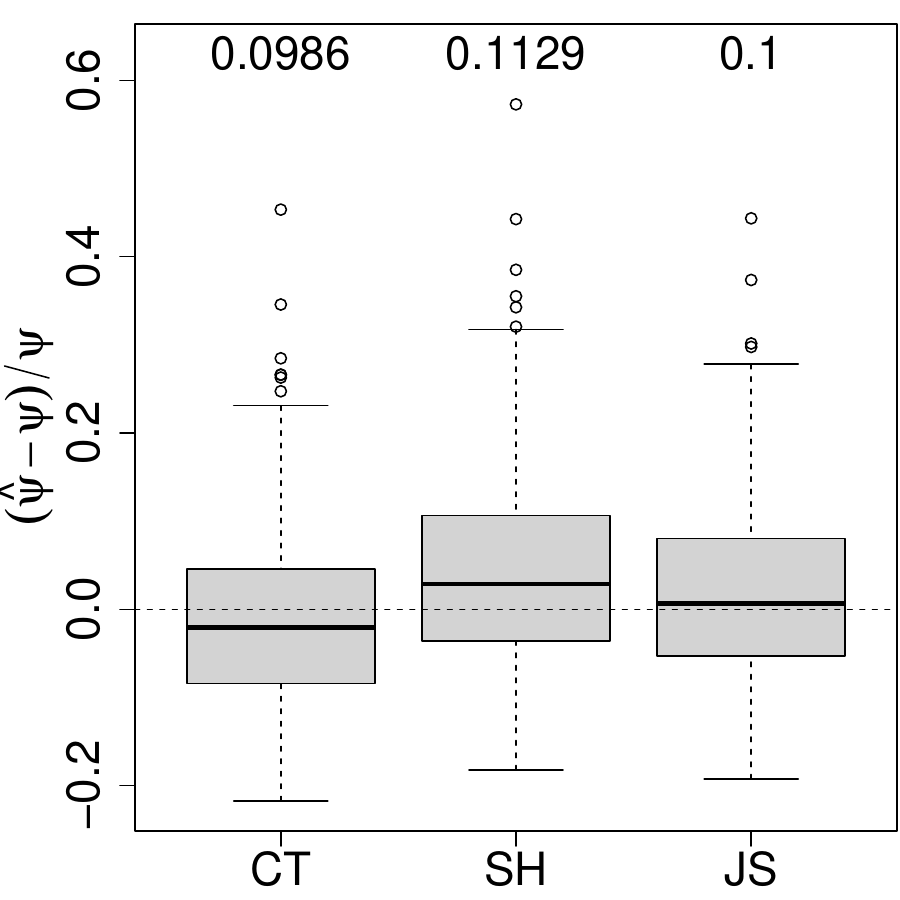} &
\includegraphics[width=0.23\textwidth]{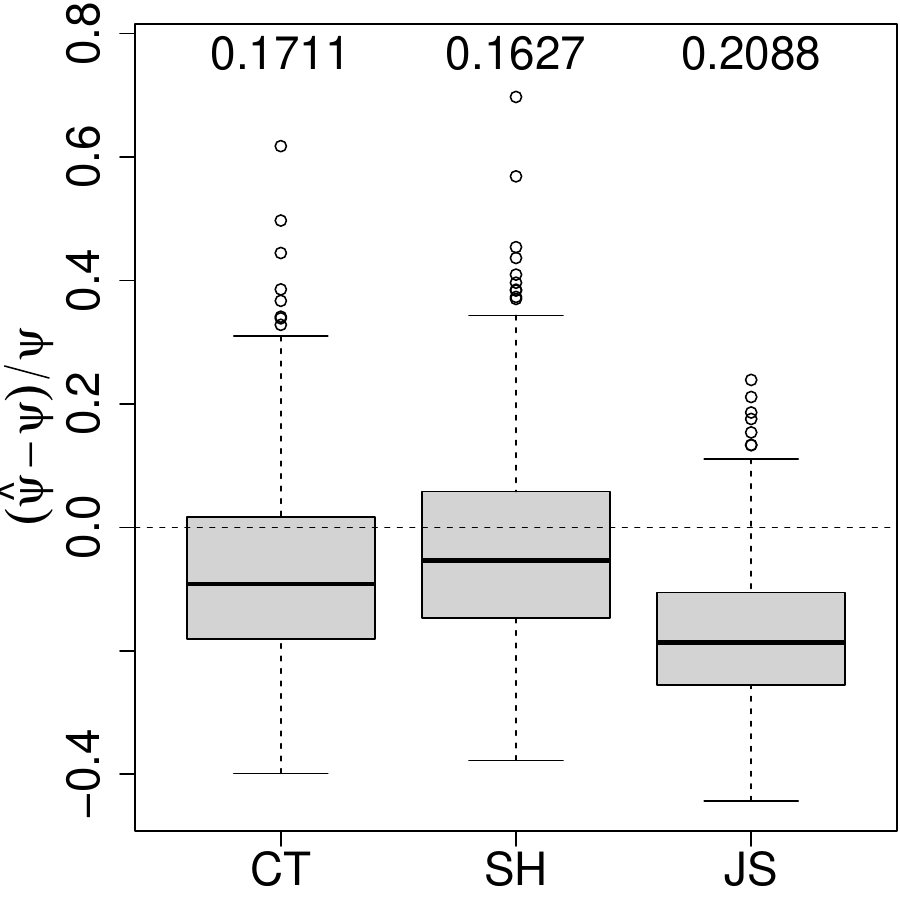} &
\includegraphics[width=0.23\textwidth]{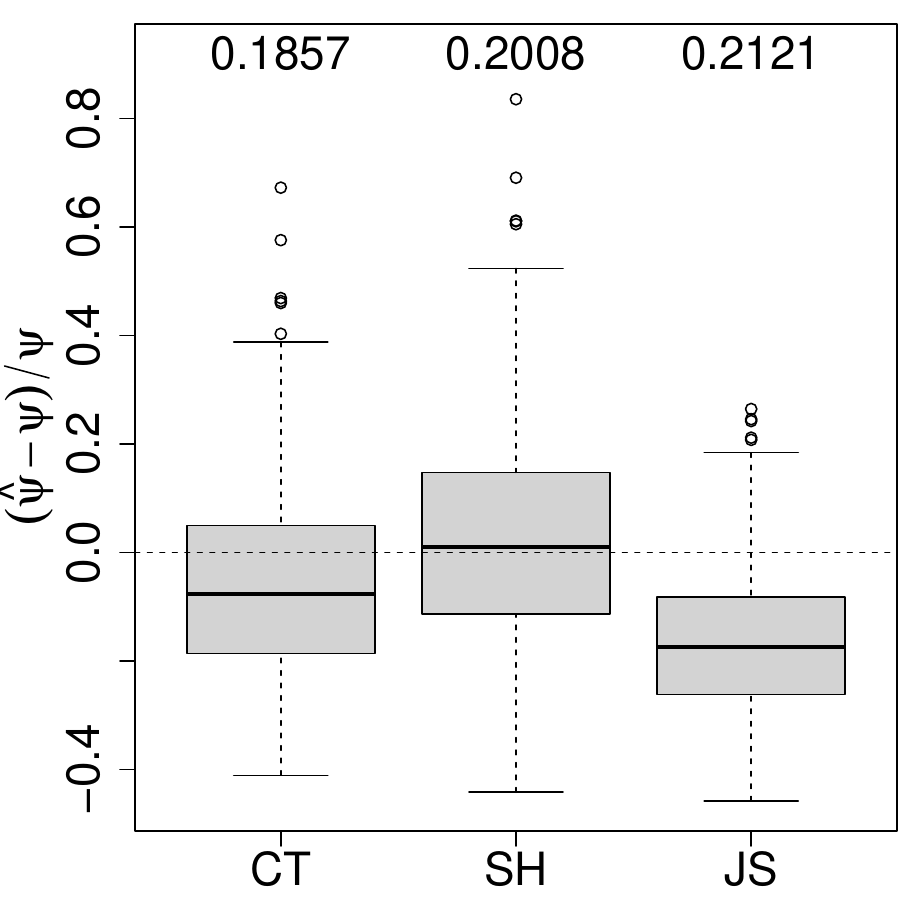}
\\
\quad Density \#5 &\quad Density \#6 &\quad Density \#7&\quad Density \#8\\
\includegraphics[width=0.23\textwidth]{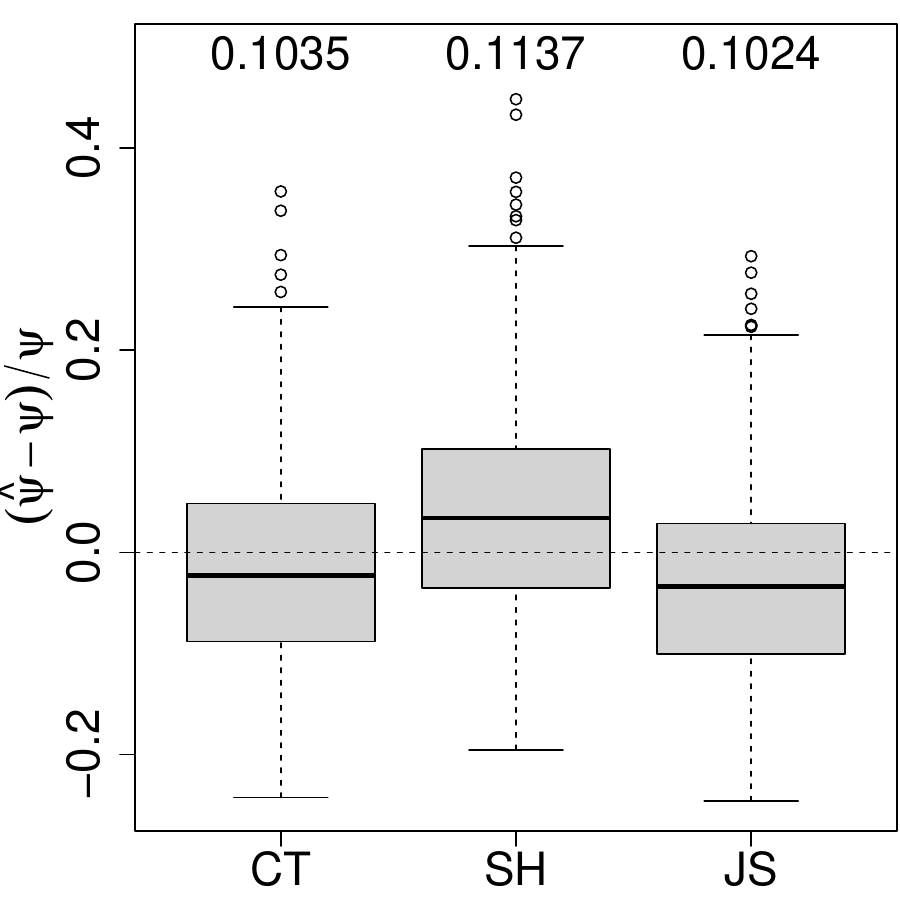} &
\includegraphics[width=0.23\textwidth]{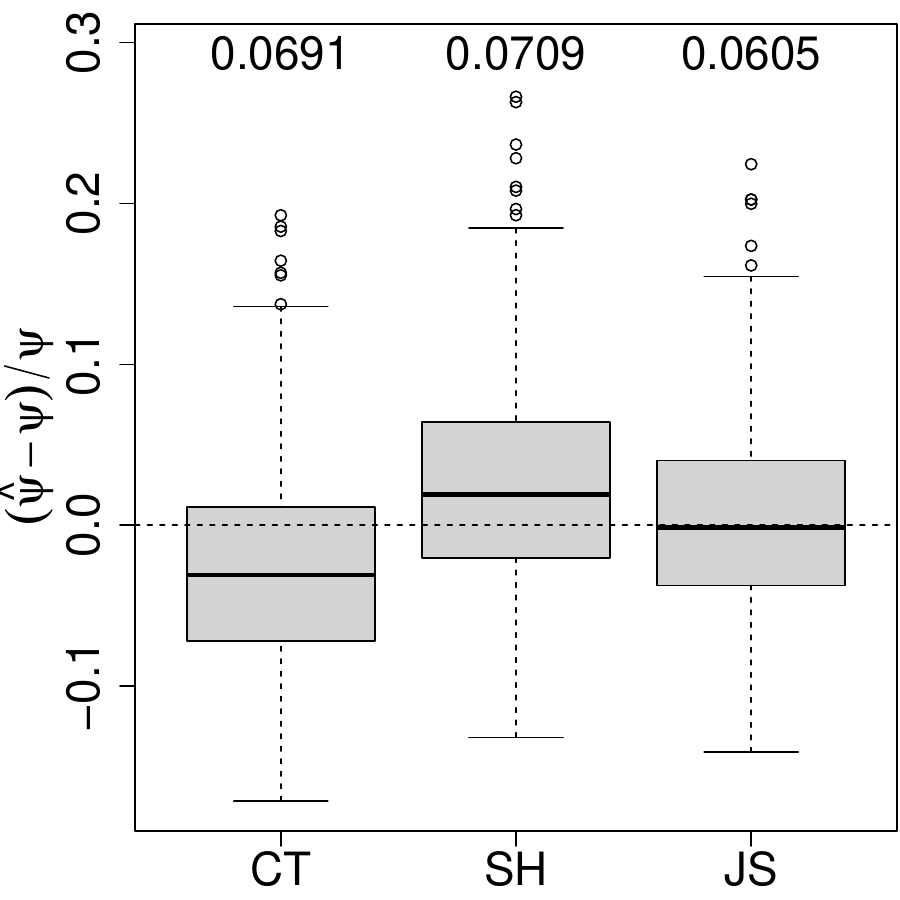} &
\includegraphics[width=0.23\textwidth]{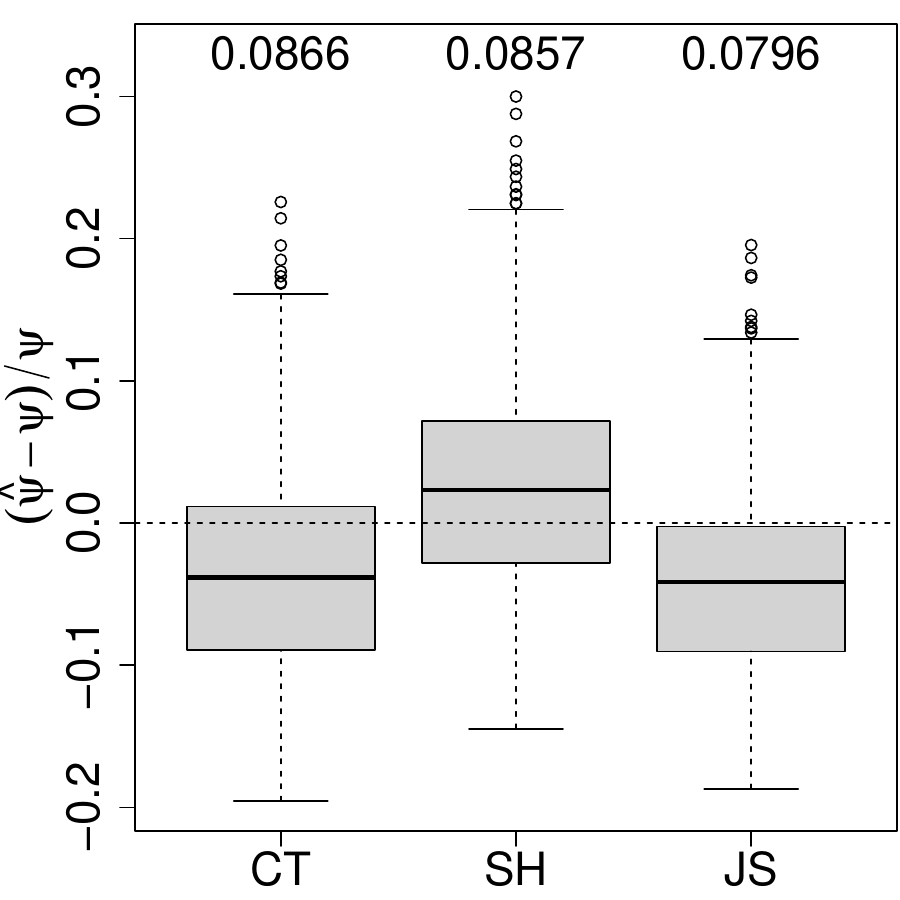} &
\includegraphics[width=0.23\textwidth]{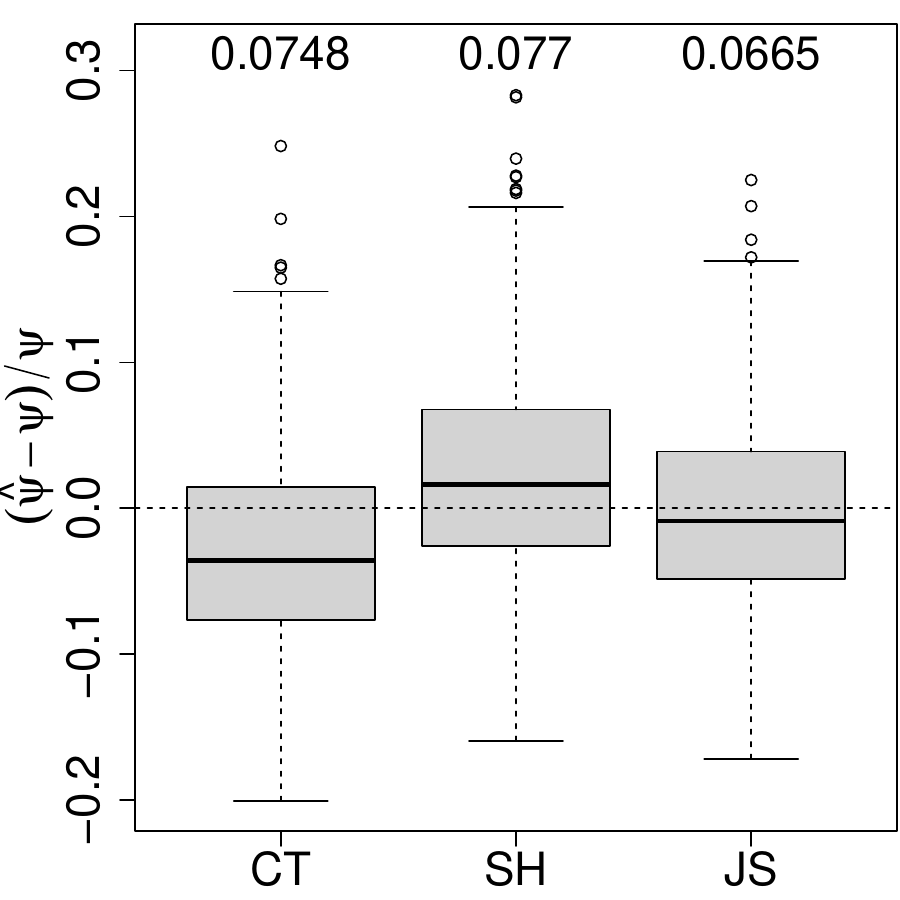} 
\\
\quad Density \#9&\quad Density \#10 &\quad Density \#11 &\quad Density \#12\\
\includegraphics[width=0.23\textwidth]{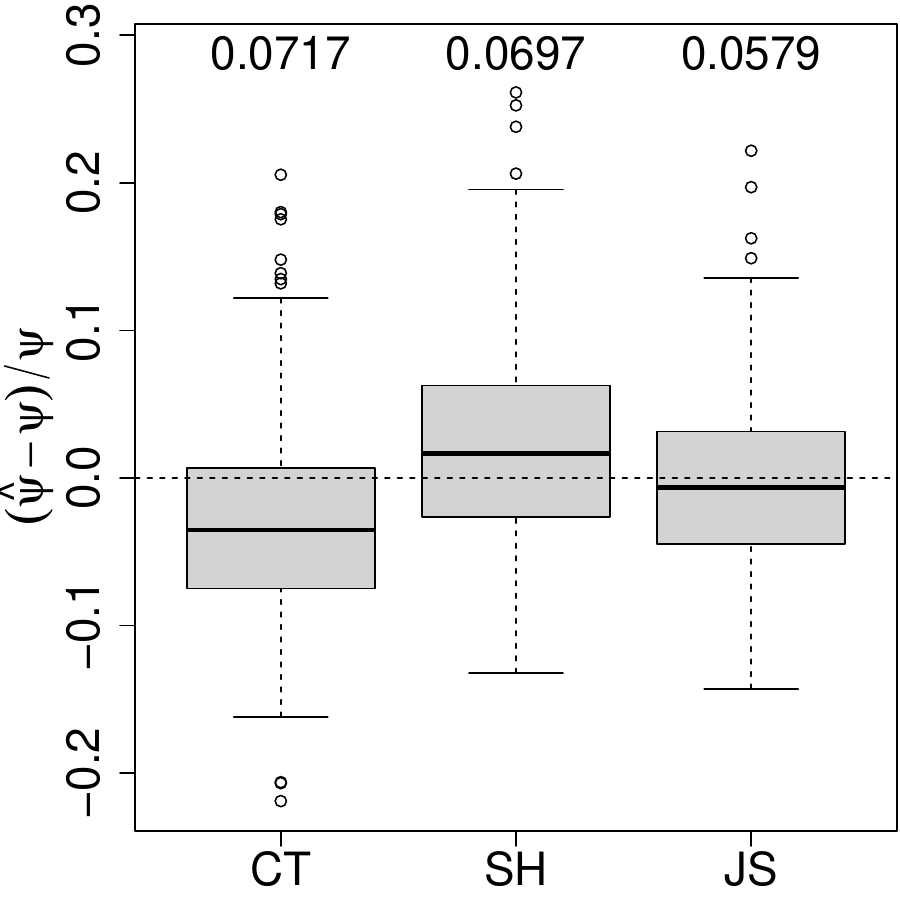}  &
\includegraphics[width=0.23\textwidth]{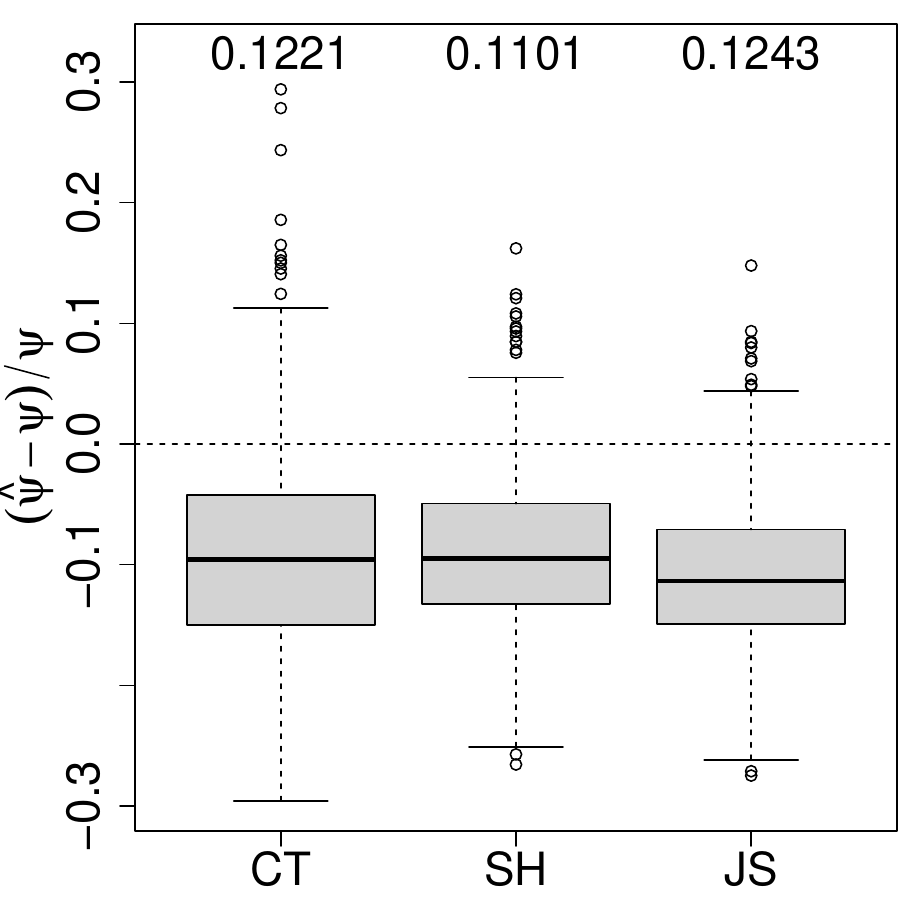} &
\includegraphics[width=0.23\textwidth]{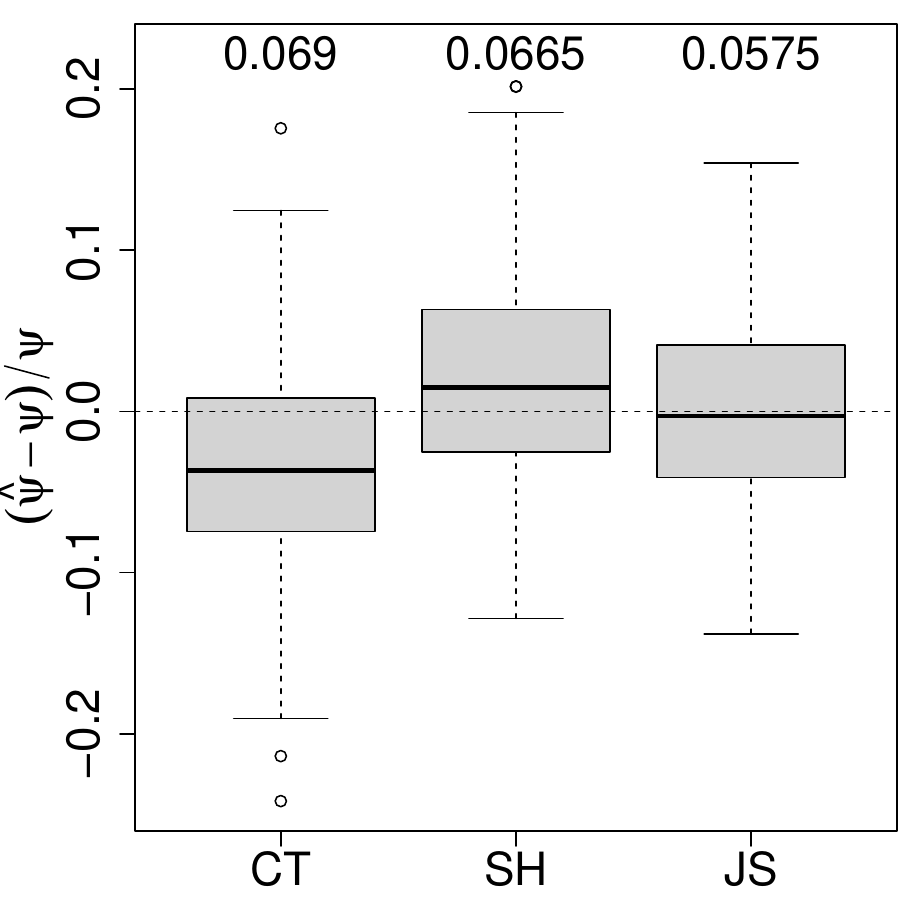} &
\includegraphics[width=0.23\textwidth]{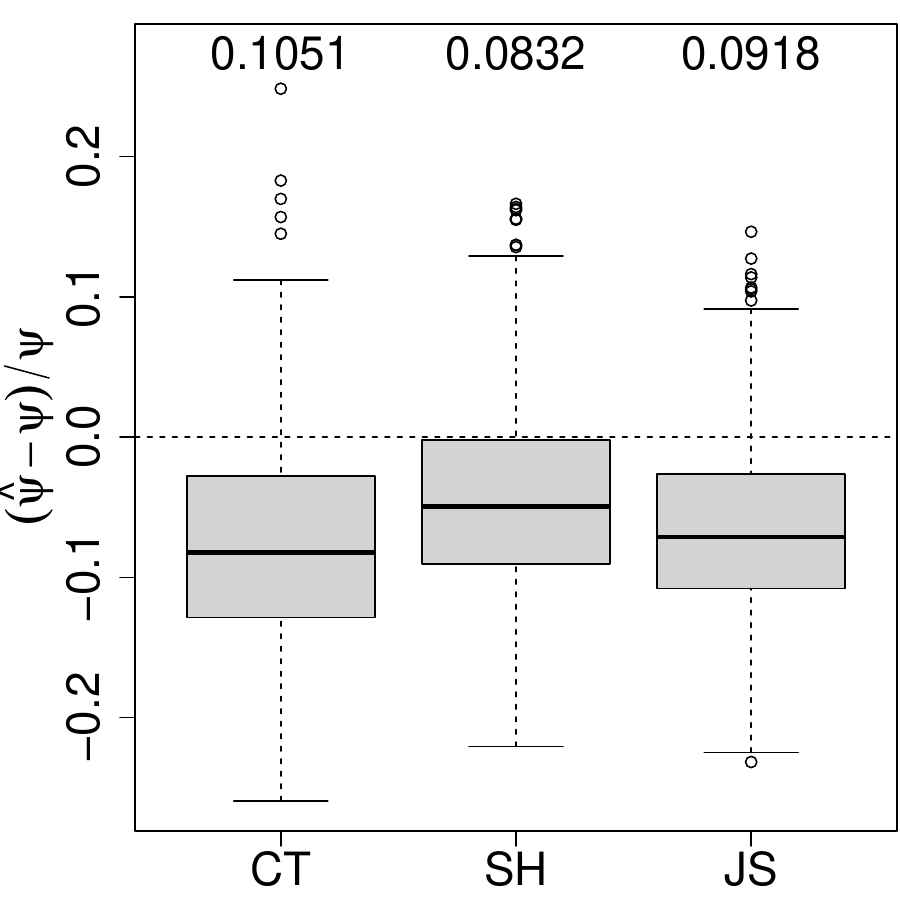}
\\
\quad Density \#13&\quad Density \#14 &\quad Density \#15 &\quad Density \#16\\
\includegraphics[width=0.23\textwidth]{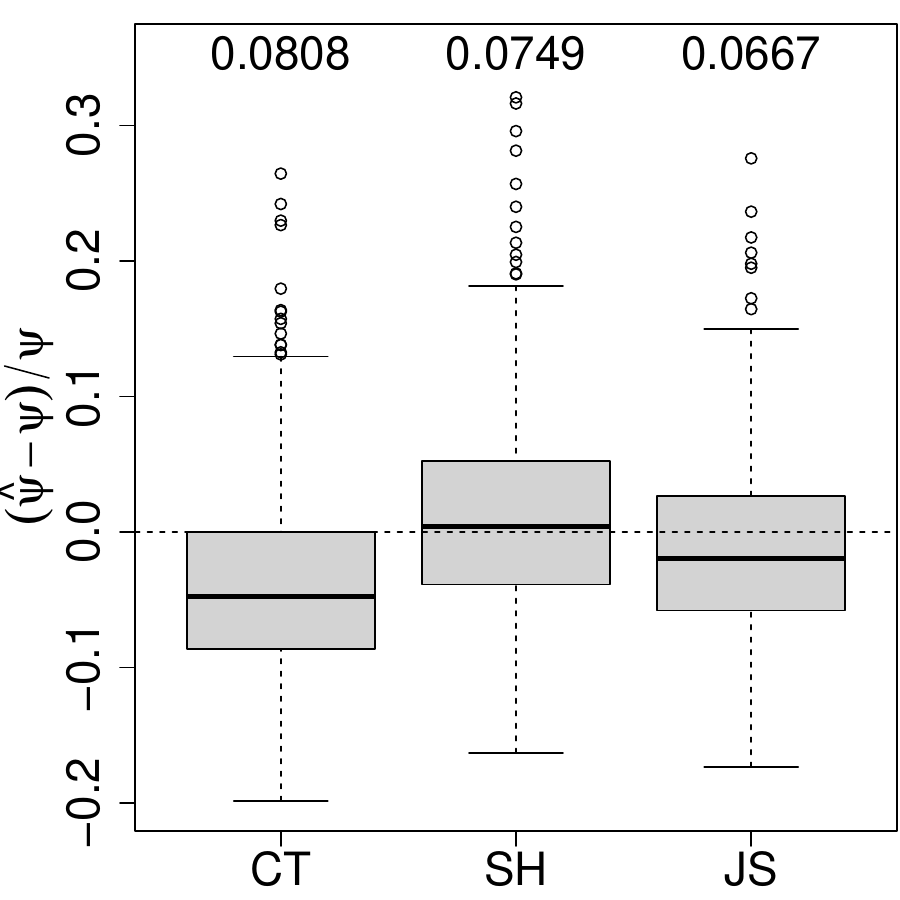} &
\includegraphics[width=0.23\textwidth]{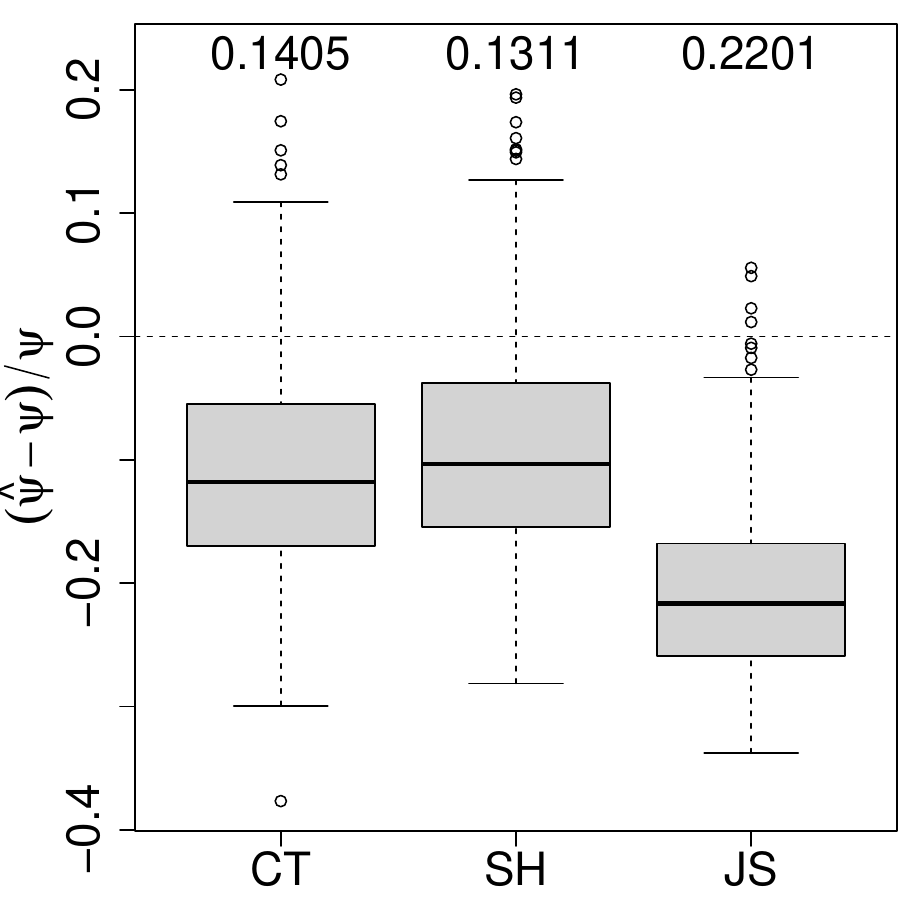} &
\includegraphics[width=0.23\textwidth]{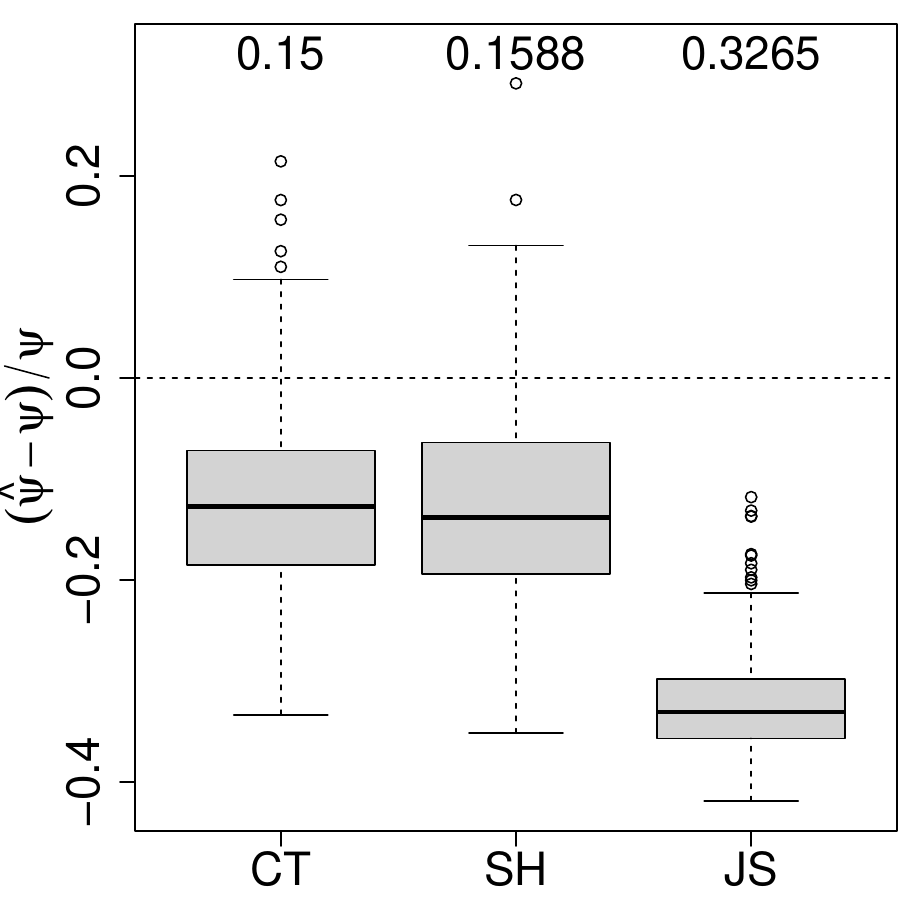} &
\includegraphics[width=0.23\textwidth]{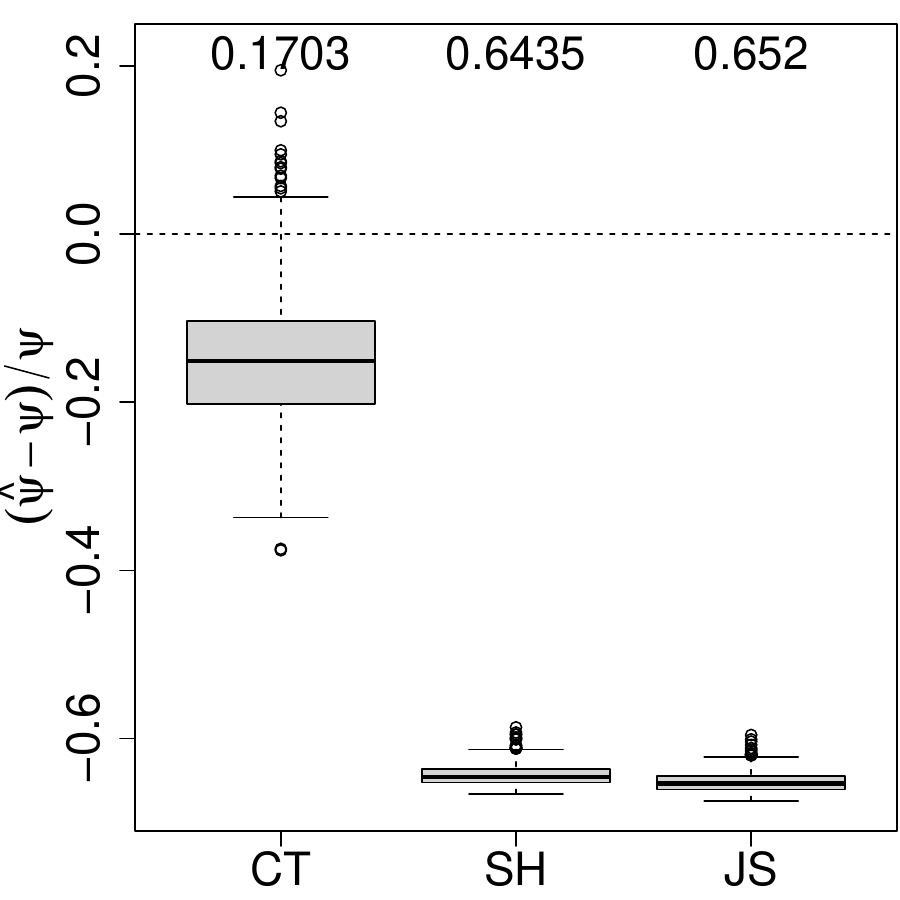}
\end{tabular}
\caption{\it Distribution boxplots for the relative error of the estimators $\widehat\psi_{\rm CT}$, $\widehat\psi_{\rm SH}$  and $\widehat\psi_{\rm JS}$ for $n=100$.}
\label{fig:box100}
\end{figure}

\begin{figure}[!pt]
\centering
\begin{tabular}{cccc}
\quad Density \#1&\quad Density \#2 &\quad Density \#3&\quad  Density \#4\\
\includegraphics[width=0.23\textwidth]{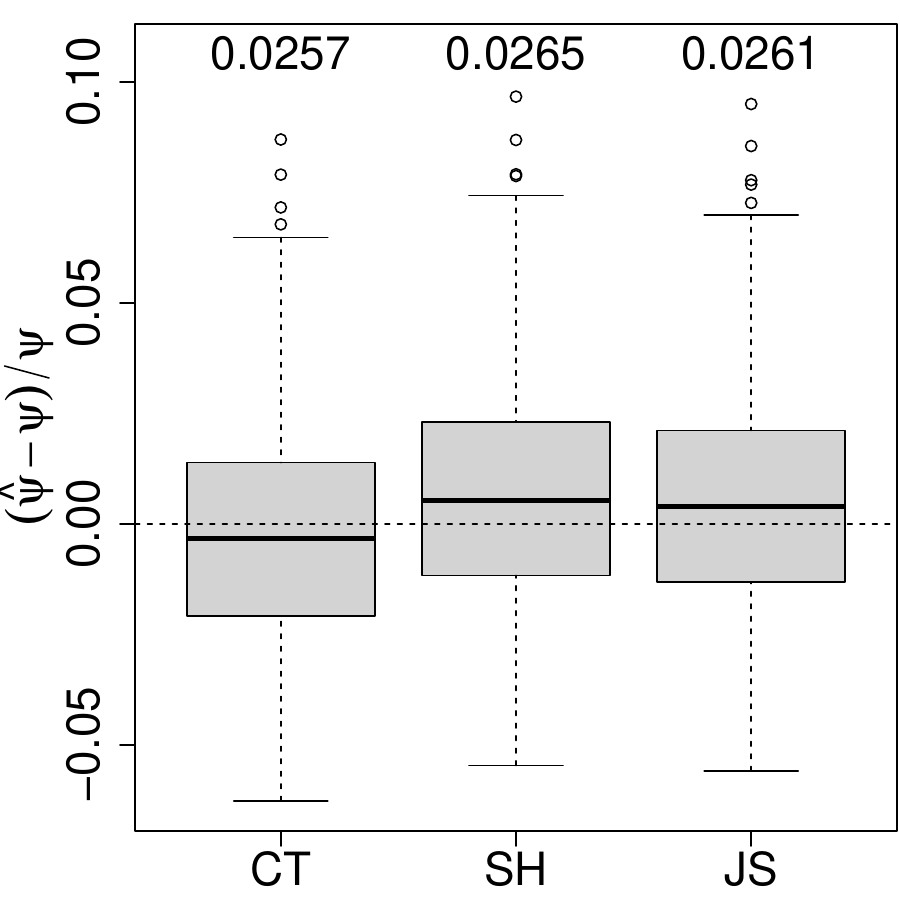} &
\includegraphics[width=0.23\textwidth]{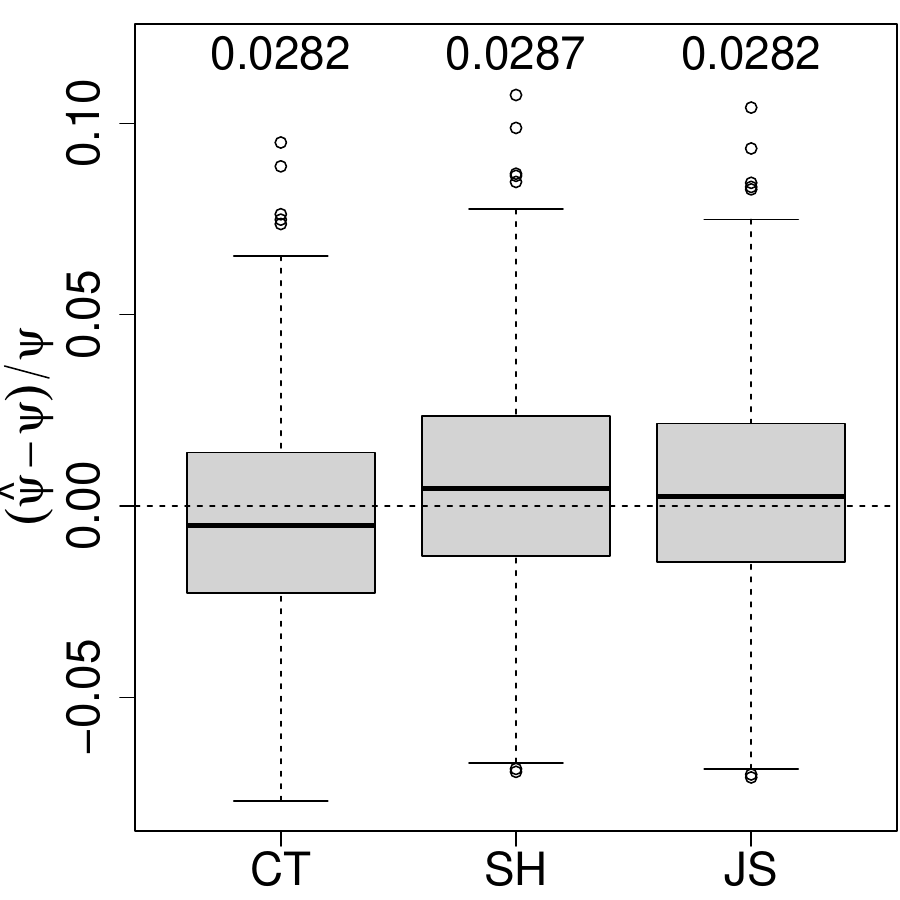} &
\includegraphics[width=0.23\textwidth]{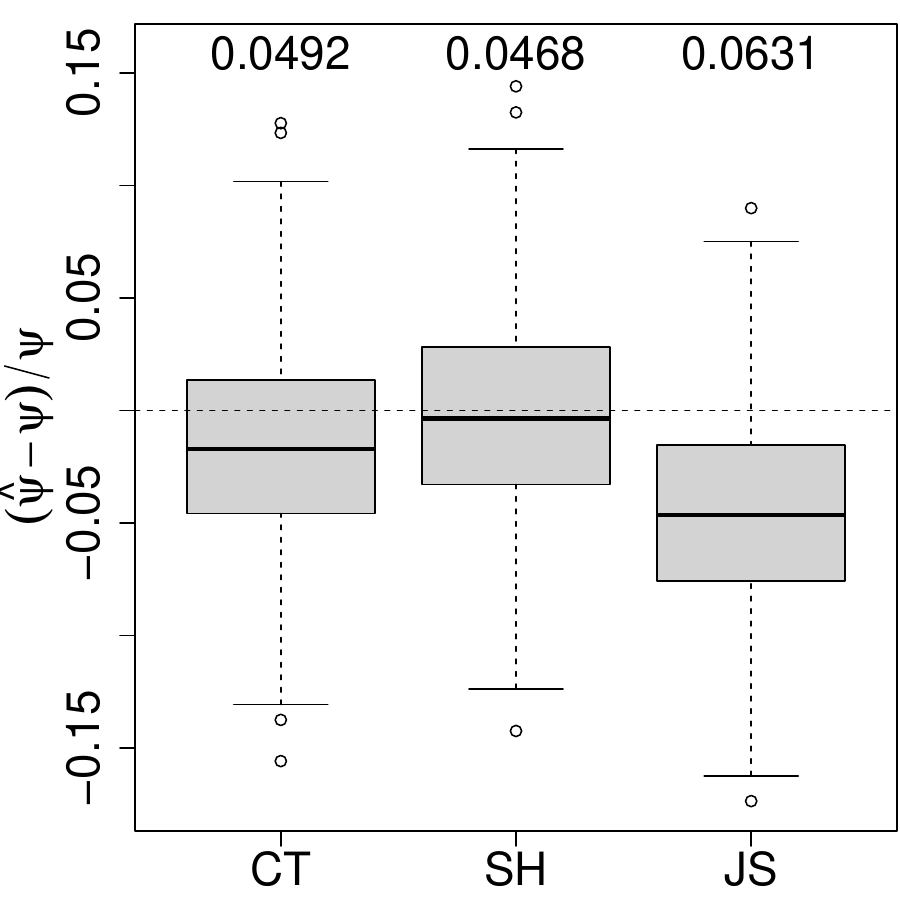} &
\includegraphics[width=0.23\textwidth]{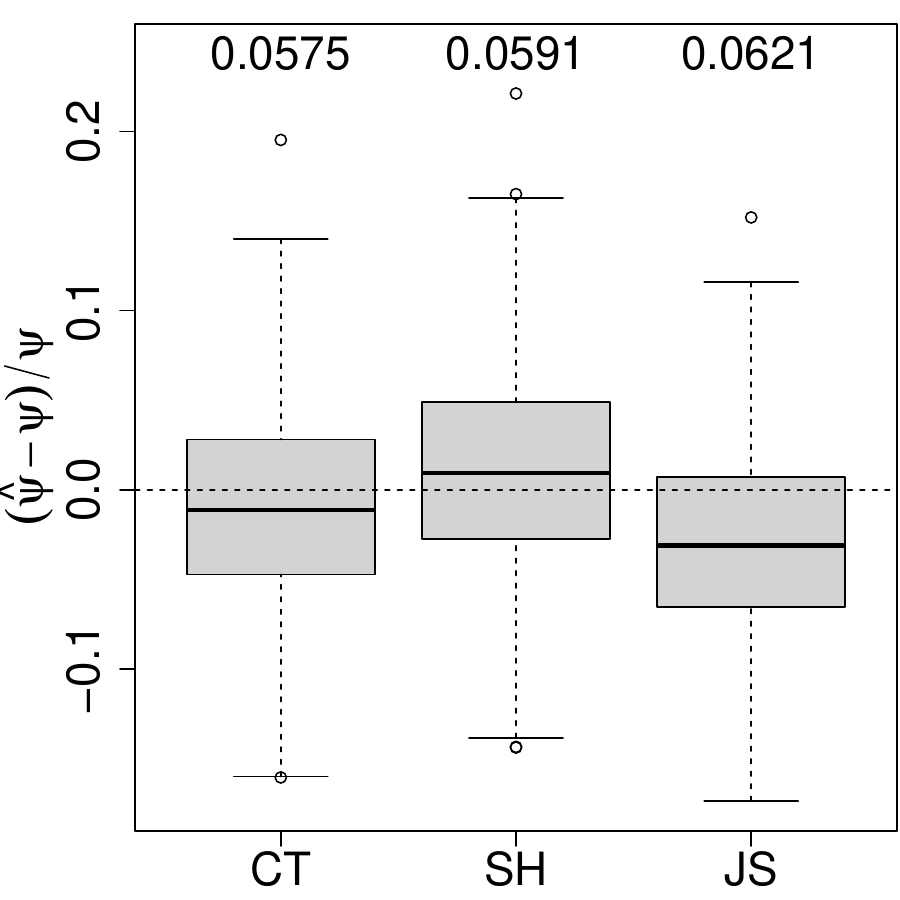}
\\
\quad Density \#5 &\quad Density \#6 &\quad Density \#7&\quad Density \#8\\
\includegraphics[width=0.23\textwidth]{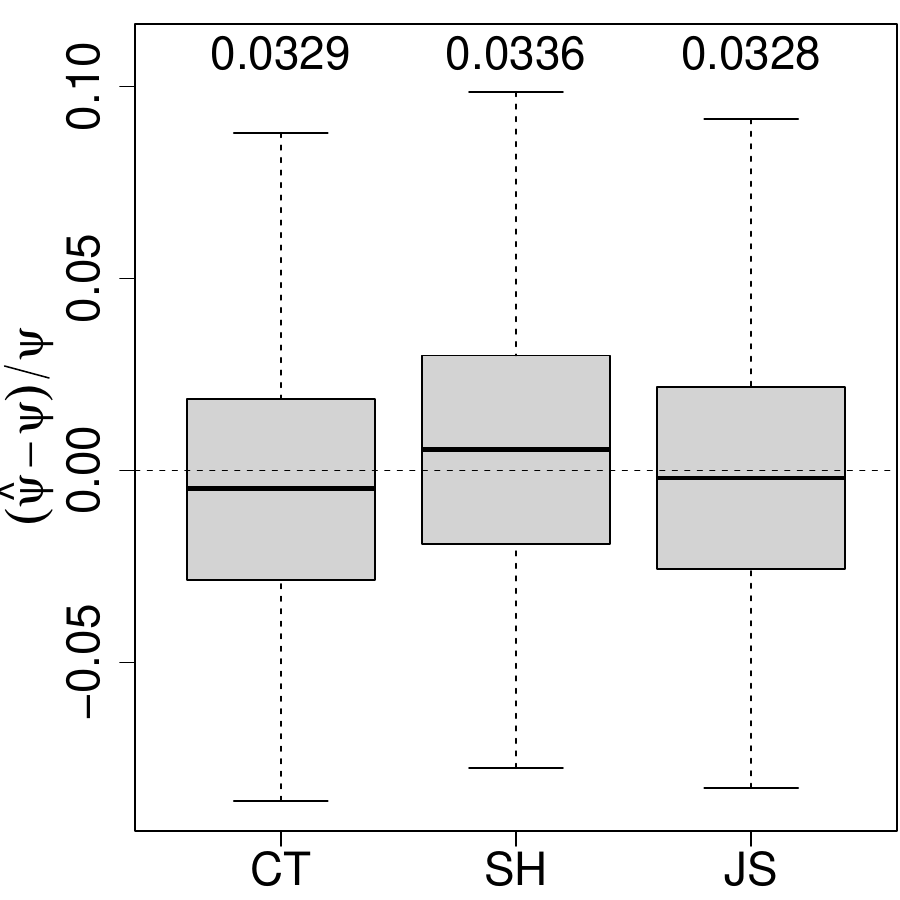} &
\includegraphics[width=0.23\textwidth]{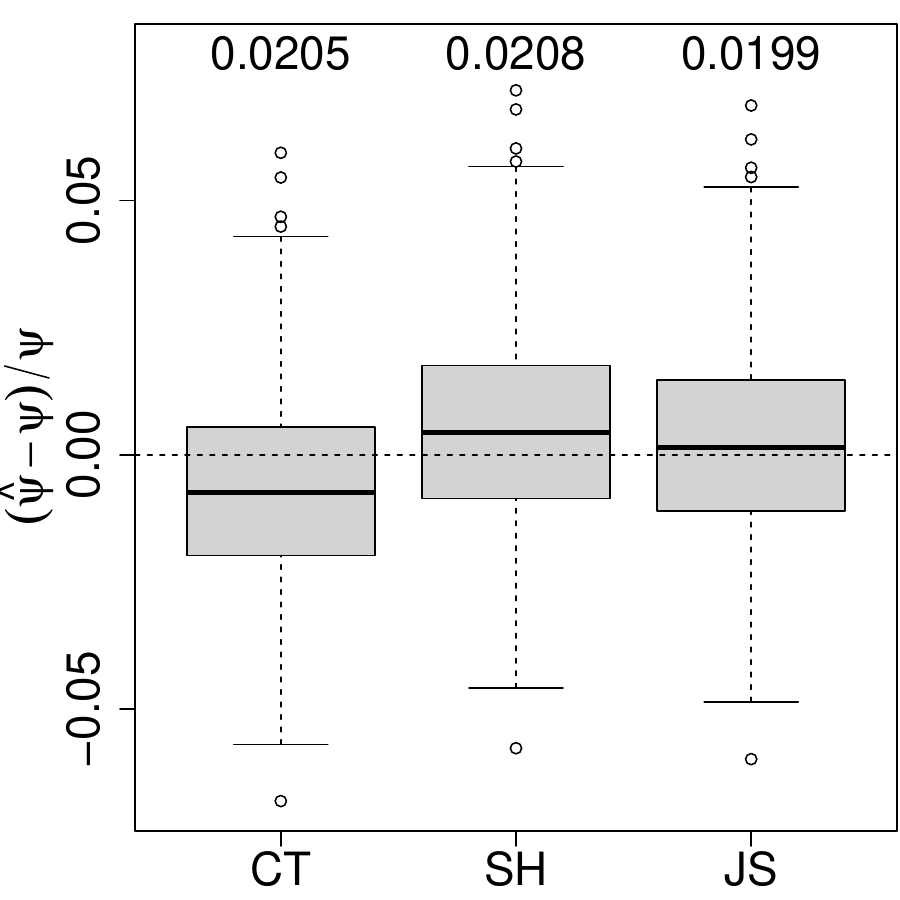} &
\includegraphics[width=0.23\textwidth]{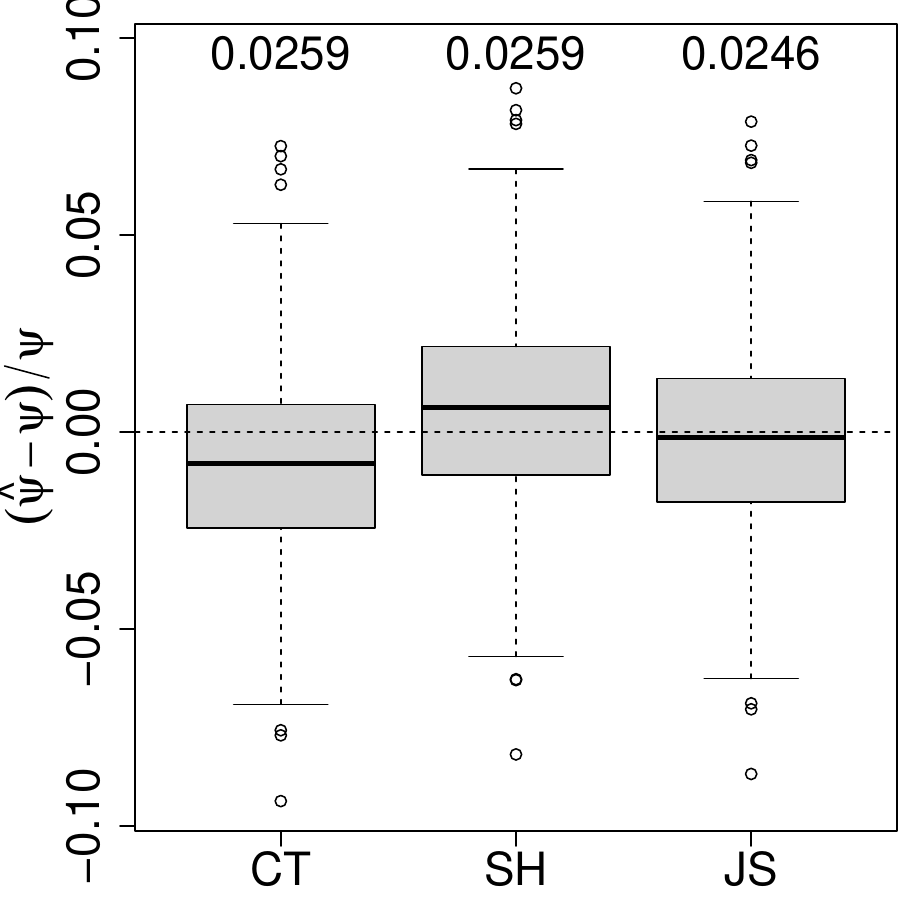} &
\includegraphics[width=0.23\textwidth]{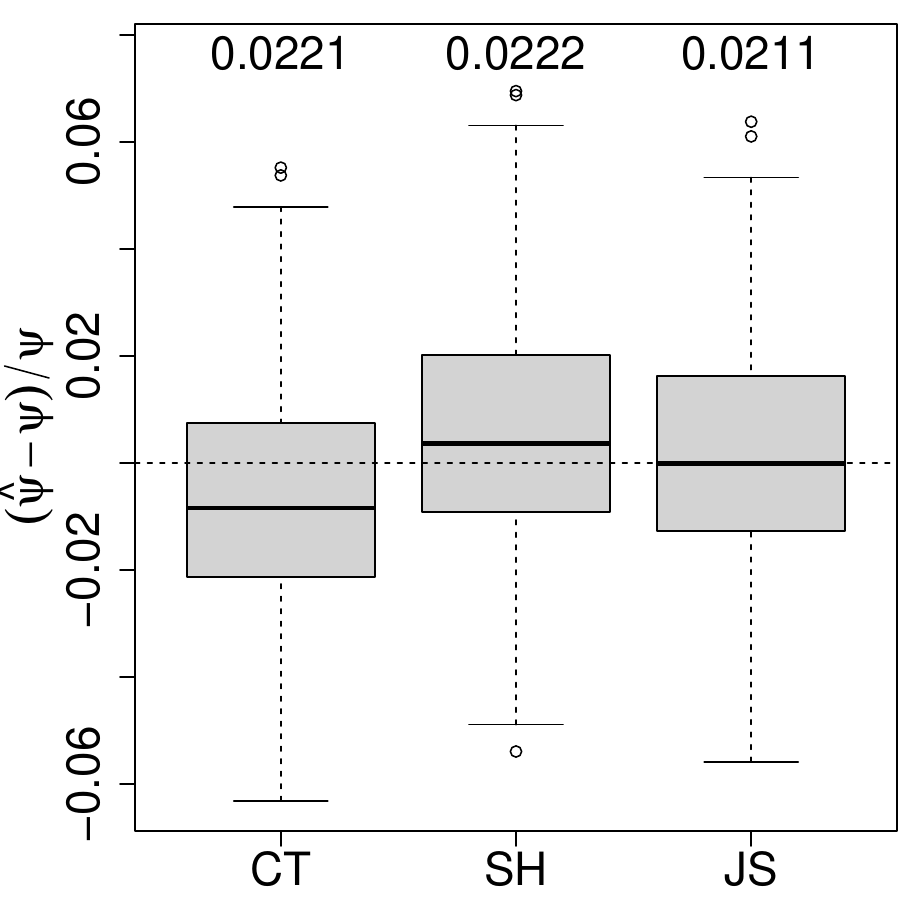} 
\\
\quad Density \#9&\quad Density \#10 &\quad Density \#11 &\quad Density \#12\\
\includegraphics[width=0.23\textwidth]{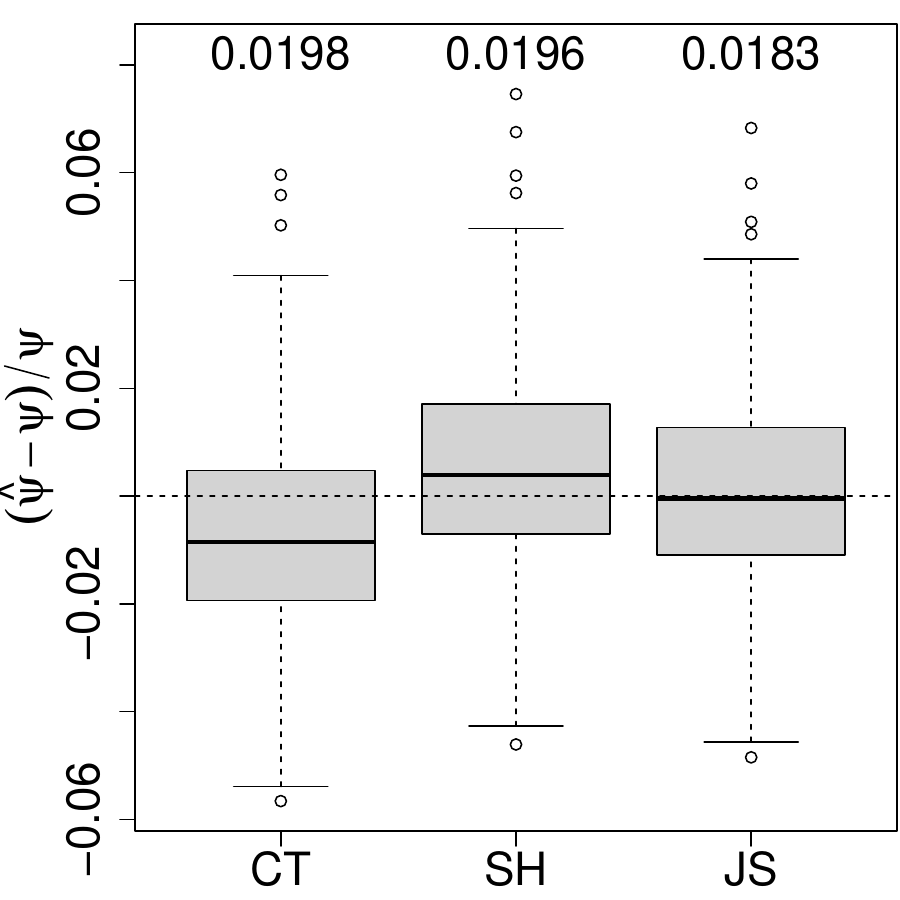}  &
\includegraphics[width=0.23\textwidth]{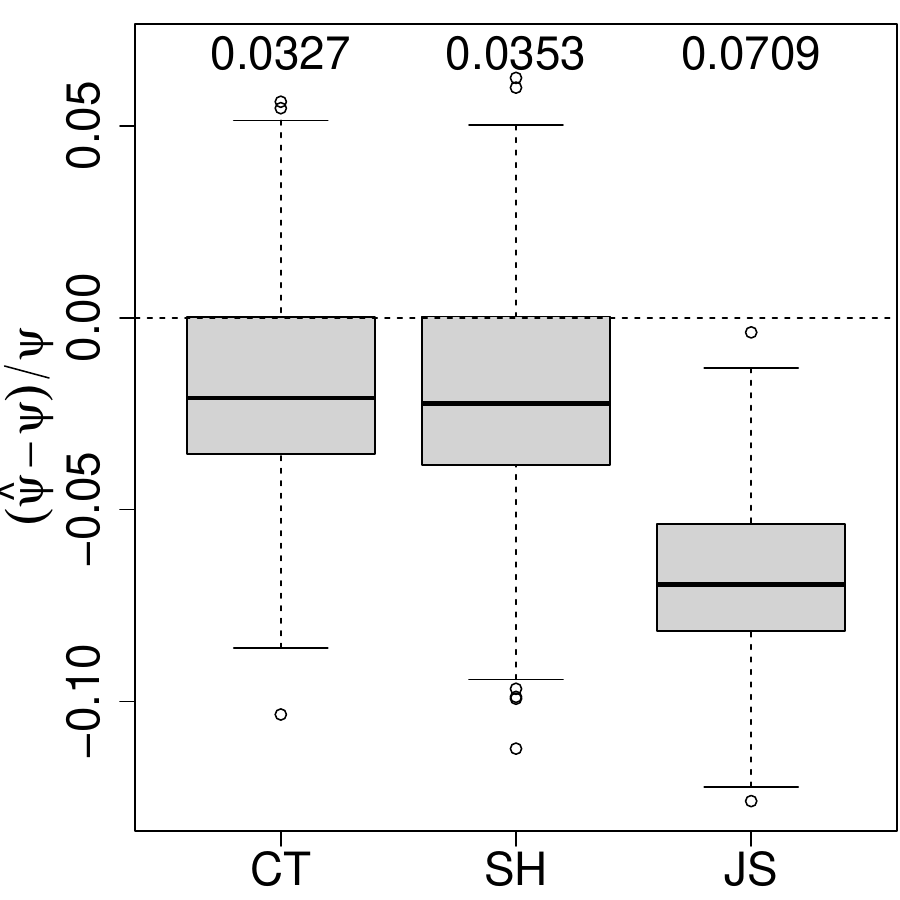} &
\includegraphics[width=0.23\textwidth]{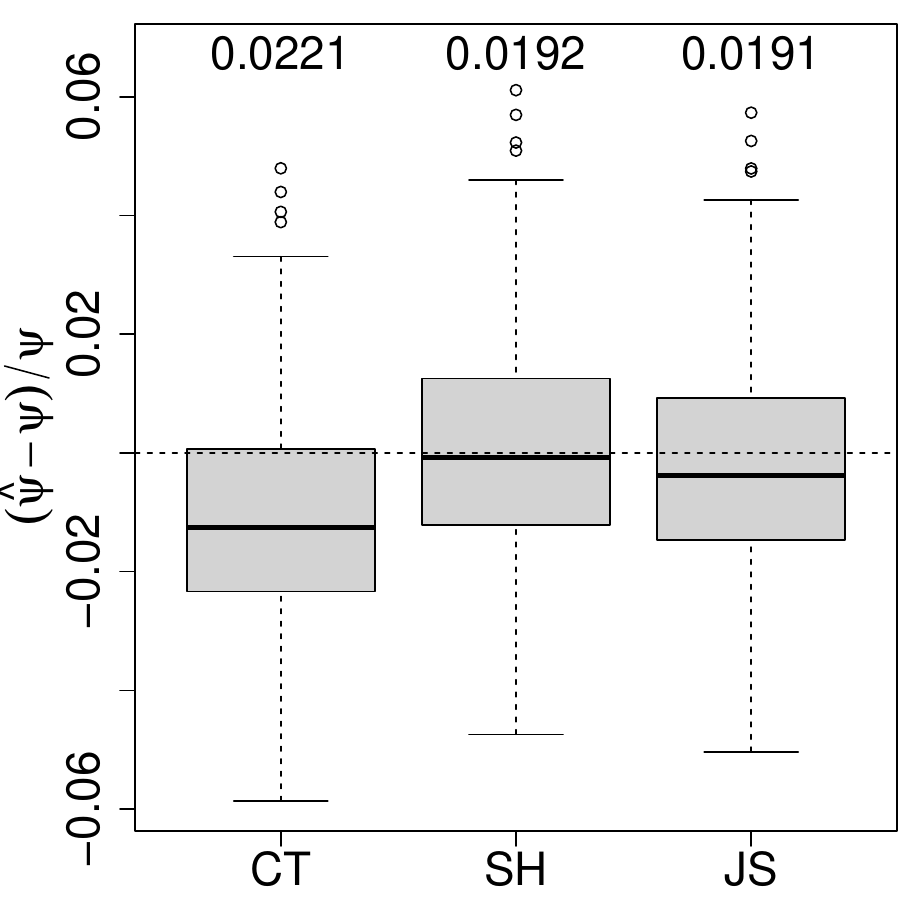} &
\includegraphics[width=0.23\textwidth]{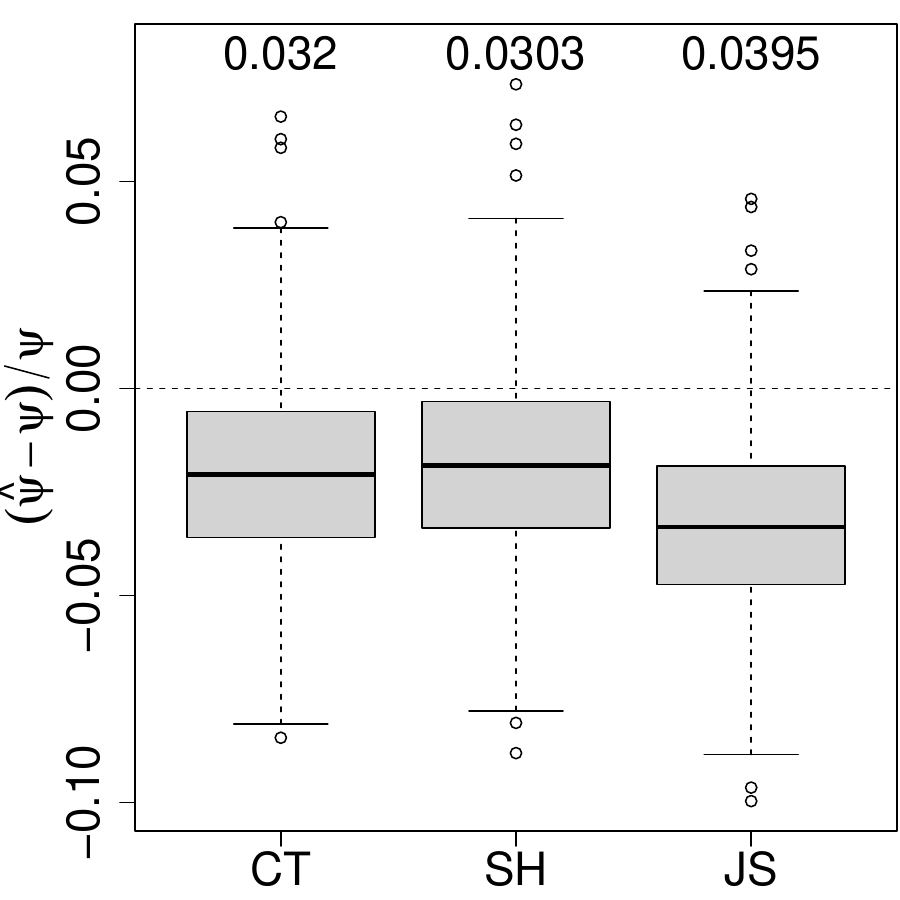}
\\
\quad Density \#13&\quad Density \#14 &\quad Density \#15 &\quad Density \#16\\
\includegraphics[width=0.23\textwidth]{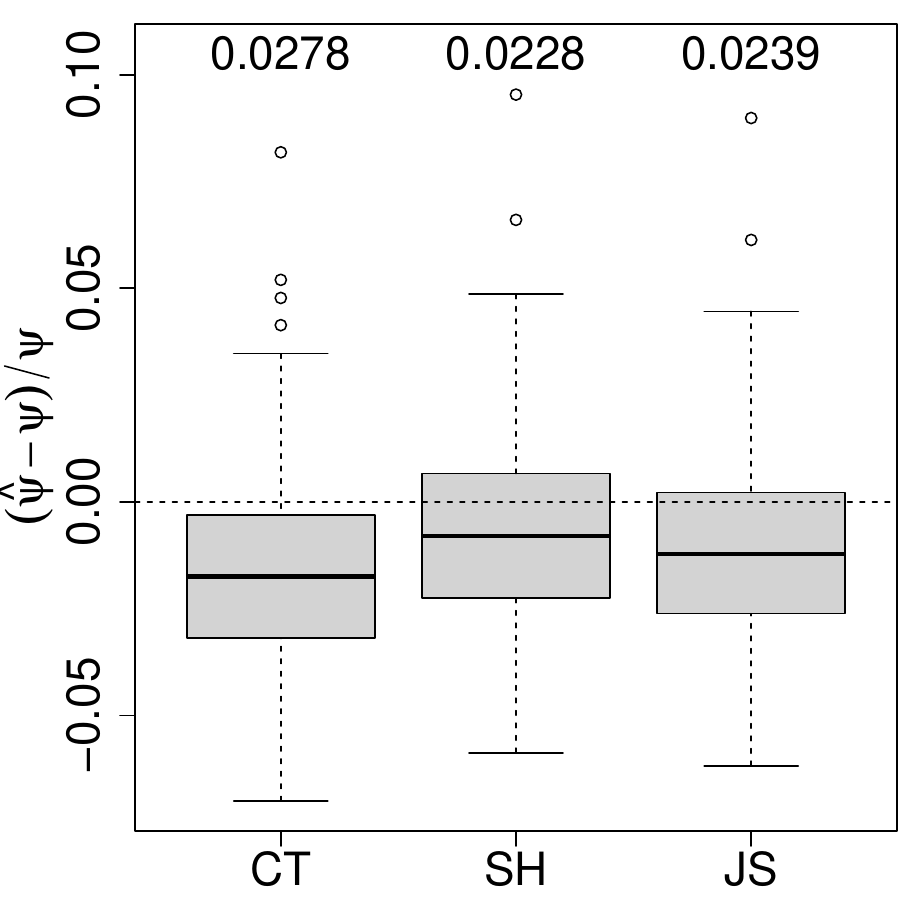} &
\includegraphics[width=0.23\textwidth]{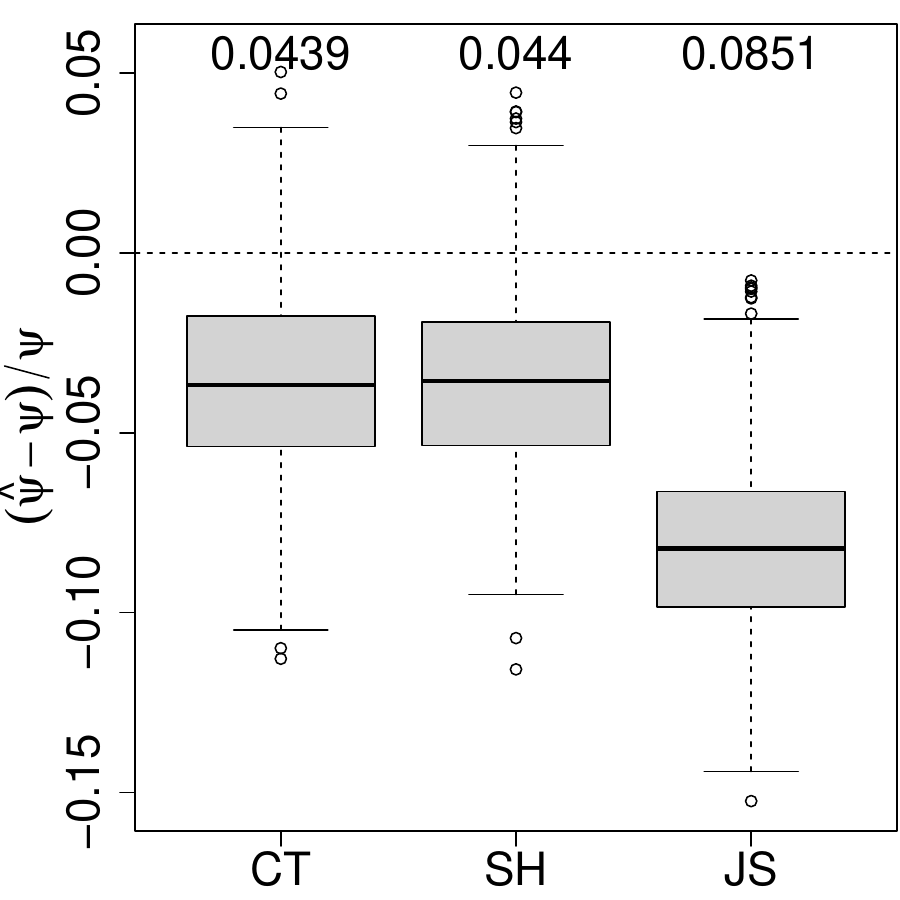} &
\includegraphics[width=0.23\textwidth]{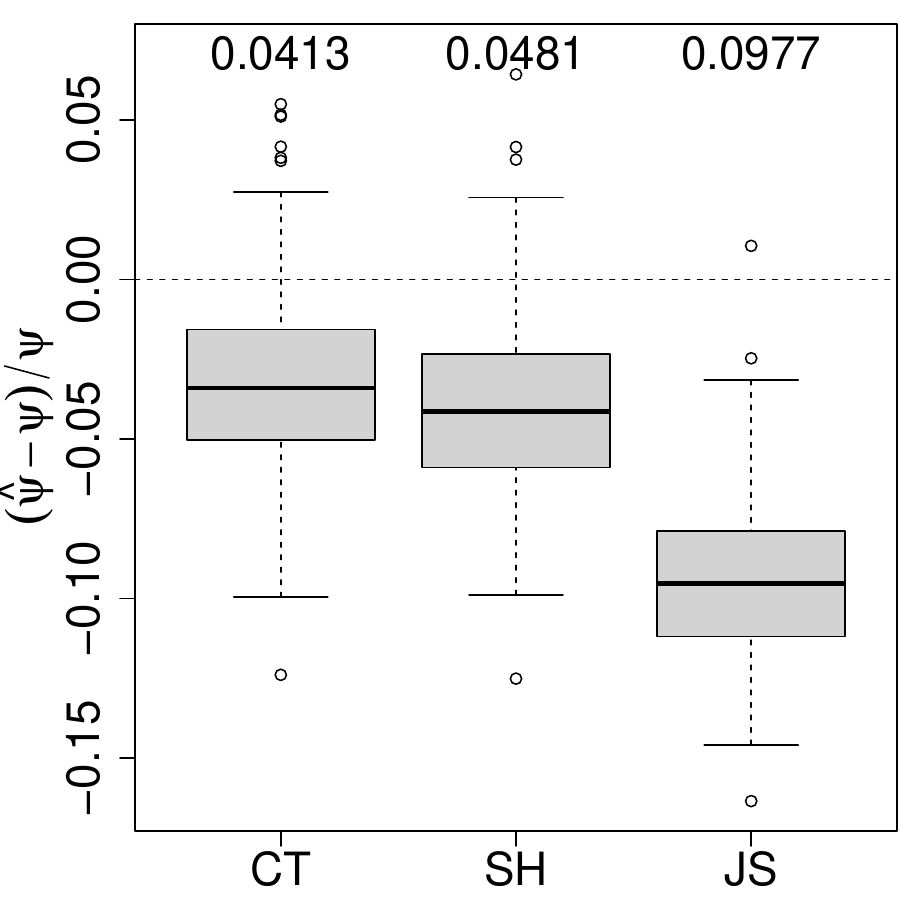} &
\includegraphics[width=0.23\textwidth]{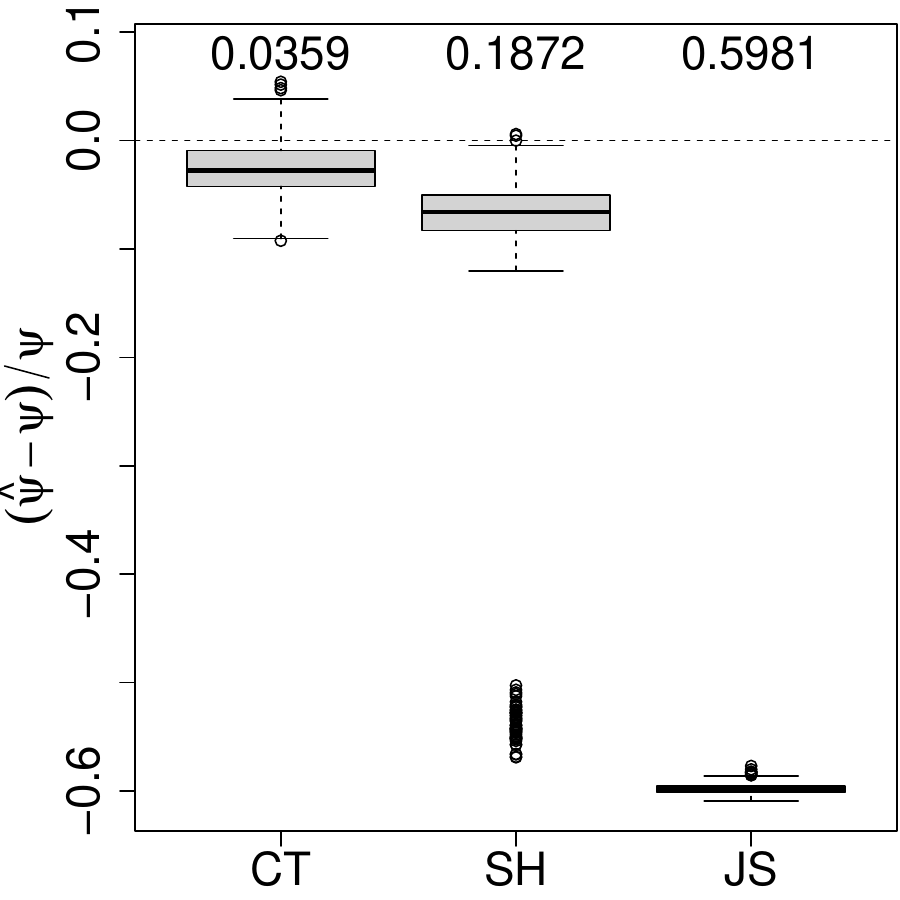}
\end{tabular}
\caption{\it Same as Figure \ref{fig:box100}, but for $n=1000$.}
\label{fig:box1000}
\end{figure}

Figures \ref{fig:box100} and \ref{fig:box1000} (for $n=100$ and $n=1000$, respectively) contain boxplots showing the distribution of the relative errors for the estimators $\widehat\psi_{\rm CT}$, $\widehat\psi_{\rm SH}$ and $\widehat\psi_{\rm JS}$ in the study, for each of the test densities considered. On the top of each boxplot there is also a number, showing the sample relative root mean squared error of the corresponding estimator for the test density under consideration, along the $B=500$ simulation runs, as defined in the main text.

These results reveal that, for the group of easy-to-estimate densities, the plug-in-type estimator $\widehat \psi_{\rm JS}$ performs quite satisfactorily. But it does not appear so competitive for some of the other density models, especially for densities 3, 4, 10, 12, 14, 15 and 16. The classical alternative $\widehat \psi_{\rm SH}$, based on a solve-the-equation rule, seems to adapt better to the more difficult-to-estimate scenarios. However, it also fails abysmally for Density \#16, more markedly with sample size $n=100$.

The breakdown of both estimators $\widehat \psi_{\rm JS}$ and $\widehat \psi_{\rm SH}$ for this model \#16 is caused by the use of a quite inadequate reference distribution at their starting steps. This is an unfortunate issue that the new estimator $\widehat \psi_{\rm CT}$ does not present, since it does not depend on the choice of a reference distribution. This feature pays off extremely well for this model, since $\widehat \psi_{\rm CT}$ appears to have a clear advantage over the two other estimators for this Density \#16. But in fact, the new proposal shows a very competitive performance along the whole group of test densities, ranking either first or very close to the first one for all density models, as reflected in the summary statistics reported in the main text.

\end{document}